\documentclass[a4paper,english,prb, showpacs]{revtex4}
\usepackage[T1]{fontenc}
\usepackage[latin1]{inputenc}
\usepackage{babel}
\usepackage{graphics}

\begin{document}

\title{Improving the convergence of defect calculations in supercells -
an \emph{ab initio} study of the neutral silicon vacancy}

\author{M~I~J~Probert%
\thanks{Author to whom correspondence should be addressed 
}}

\affiliation{Department of Physics, University of York, Heslington, York, YO10
5DD, U.K.}

\author{M~C~Payne}

\affiliation{Cavendish Laboratory, University of Cambridge, Madingley Road, Cambridge,
CB3 0HE, U.K.}

\date{\today{}}

\begin{abstract}
We present a systematic methodology for the accurate calculation of
defect structures in supercells which we illustrate with a study of
the neutral vacancy in silicon. This is a prototypical defect which
has been studied extensively using \emph{ab initio} methods, yet remarkably
there is still no consensus about the energy or structure of this
defect, or even whether the nearest neighbour atoms relax inwards
or outwards. In this paper we show that the differences between previous
calculations can be attributed to supercell convergence errors, and
we demonstrate how to systematically reduce each such source of error.
The various sources of scatter in previous theoretical studies are
discussed and a new effect, that of supercell symmetry, is identified.
It is shown that a consistent treatment of this effect is crucial
to understanding the systematic effects of increasing the supercell
size. This work therefore also presents the best converged \emph{ab
initio} study of the neutral silicon vacancy to date.
\end{abstract}


\maketitle

\section{Introduction}

\label{sec:intro} There has been much interest over the last 10 years
in the calculation of the properties of defects in solids. Various
theoretical techniques can be used, but two of the most common are
\emph{ab initio} electronic structure calculations of the defect in
either a periodic supercell or a cluster. Both of these techniques
have advantages and disadvantages. Periodic boundary conditions are
a natural representation of a crystal, but the introduction of a defect
into the supercell results in the calculation of a periodic array
of defects and not an isolated defect. A cluster calculation might
therefore seem more appropriate, but this introduces different problems
due to finite size effects and the possibility of interaction between
the defect and the surface of the cluster. In this paper, we will
focus exclusively on the periodic supercell technique.

There are various technical problems that must be overcome before
a supercell calculation becomes an accurate representation of an isolated
defect in bulk material. Many of these are already known, but not
all are widely appreciated. It is the aim of this paper to synthesize
{}``best-practice'' into a systematic approach to the study of defects
in periodic supercells and show how best to overcome all these problems.
As an example of this methodology, we shall consider the neutral vacancy
in silicon, which is perhaps the simplest example of a point defect
in a crystal. It is certainly one that has been often studied experimentally
and theoretically. However, whilst the experimental picture is reasonably
clear, there are some properties of the defect, such as its local
volume, which cannot be extracted from experimental data and for which
reliable thoeretical values would be very welcome. Unfortunately,
there is some confusion about the results of theoretical studies which
has prevented definitive statements being made about these properties.
Recent work has achieved some degree of consensus, but there is still
a lot of scatter in the results which is not well understood. The
aim of this work is to show how a systematic approach to the study
of defects can explain the origin of the scatter in earlier works
on the neutral silicon vacancy, although the approach can obviously
be applied to the study of any defect. By reducing all the systematic
errors to an acceptable level using this methodology, we therefore
provide the most highly converged \emph{ab initio} study of the neutral
silicon vacancy to date. 

We shall perform all our calculations using Density Functional Theory
(see ref. \onlinecite{JonesG89} for a review) but the general features
of our methodology will be applicable to any \emph{ab initio} simulation
technique that uses periodic supercells. In this particular case we
used plane-wave pseudopotential-based DFT, which has been shown in
many previous studies to be a reliable technique for calculating many
structural properties - typically agreeing with experiment to at least
1\% accuracy (see ref. \onlinecite{PayneTAAJ92} for a review of total
energy calculations using this technique). 

More accurate treatments of the electronic structure exist, such as
Quantum Monte Carlo (see ref. \onlinecite{FoulkesMNR01} for a review),
but these are much more expensive to apply and are not yet routinely
used for such calculations. One reason for this is that whilst QMC
is currently {}``state-of-the-art'' in terms of its accuracy in
calculating the electronic structure, it is unable to calculate forces
and is therefore not a practical methodology for structure determination.
There is therefore still a role for DFT calculations to determine
structural properties, as well as being a source of trial input wavefunctions
to QMC calculations at the optimal structure.

Ultimately, the results obtained will be limited by the approximations
inherent in the \emph{ab initio} technique used. In the case of the
neutral silicon vacancy calculation detailed herein, this will include
effects due to the choice of pseudopotential, choice of exchange-correlation
functional, and the neglect of zero-point motion and thermal effects.
However, if these systematic effects are to be quantified, it is important
that they are not obscured by random noise arising from other sources
of error that can be removed, such as the various kinds of supercell
convergence error that are discussed in this paper. One simple way
to estimate these effects would be to compare the DFT and QMC energies
at the same defect structure (e.g. the DFT optimal structure). However,
the supercell convergence errors discussed herein will still effect
the QMC calculation, and so need to be understood and minimised before
the comparison of two different \emph{ab initio} methodologies becomes
meaningful. The aim of this paper is to show how to systematically
reduce the different supercell convergence errors independent of the
\emph{ab initio} methodology chosen.

This paper is structured as follows: in section \ref{sec:review}
we will review what is already known about the vacancy in silicon,
in section \ref{sec: method} we will explain the key features of
our methodology in some detail, illustrated with the neutral silicon
vacancy calculation. We will report our results for the neutral silicon
vacancy in section \ref{sec:results}, and will briefly summarize
in section \ref{sec:conc}.

\section{Review}

\label{sec:review} The vacancy in silicon is a technologically important
defect, as it is known to play an important role in both self- and
impurity diffusion, and hence it is essential to have a detailed understanding
of both the electronic and ionic structure of the defect. The vacancy
also occurs in a variety of charge states, conventionally referred
to as \( V^{2+} \), \( V^{+} \), \( V^{0} \), \( V^{-} \)and \( V^{2-} \).
It is known that this system shows the \emph{negative-U} effect, that
is, \( V^{2+} \) spontaneously converts directly to \( V^{0} \).
For simplicity, this theoretical work will only focus on the neutral
vacancy, \( V^{0} \), although in some experimental techniques it
is the charged vacancies that are actually studied.

\subsection{Experimental studies}

The experimental studies have been reviewed by Watkins.\cite{Watkins86}
In summary, electron paramagnetic resonance (EPR) studies can be used
to give the symmetry and spatial distribution of the highest unpaired
localized electron state. This has shown that the symmetry of the
single neutral vacancy \( V^{0} \) is \( D_{2d} \). \cite{Watkins76}
This is understood to be due to the four dangling bonds, created by
the removal of a silicon atom from a perfect lattice, hybridizing
with each other to form two new levels. These are the \( A_{1} \)
singlet which lies deep in the bulk valence bands and the \( T_{2} \)
triplet which lies in the energy gap. The neutral vacancy has only
one of the gap states occupied, which results in a Jahn-Teller distortion,
with the ionic relaxation lowering the \( T_{d} \)-point symmetry
of the perfect lattice to that observed in the experiments.

Electron-nuclear double resonance (ENDOR) has also been used to study
the charged vacancies,\cite{SprengerMA83,SprengerMSA87} which in
general have lower symmetry than the neutral vacancy considered here.
Deep level transient spectroscopy (DLTS) has also been used to give
information about the ionization levels associated with charge state
changes.\cite{Samara87,Samara88,Samara89a,Samara89b} Positron lifetime
measurements have also given information about the defect volume associated
with charge state changes.\cite{MakinenCHMP89,MakinenHC92} However,
none of these techniques gives information on the defect volume or
formation energy of the \( V^{0} \) state. This has lead to some
confusion, with some theoretical studies claiming an outwards relaxation
of the atoms surround the vacancy, and others an inwards relaxation.

\subsection{Theoretical studies}

There have been numerous theoretical studies of the silicon vacancy
using different theoretical techniques. For example, Green's-function
calculations\cite{Scheffler82,GunnarssonJA83,SchefflerV85} predicted
an outwards relaxation of the vacancy, whilst more recent tight-binding\cite{WangCH91,MunroW99}
and \emph{ab initio} studies\cite{AntonelliB89,SuginoO92,BlochlSCLAP93,RamamoorthyP96,PankratovHDM97,ZywietzFB97,PuskaPPN98}
have proposed an inwards relaxation. Recent \emph{ab initio} cluster
calculations\cite{OgutKC97} have suggested an inwards relaxed \( D_{2d} \)
symmetry structure, whereas successive \emph{ab initio} supercell
calculations using different supercell sizes (from 32 to 216 atoms)
and different special \textbf{k}-point sampling techniques have yielded
a broad spread of formation energies(from \( 2.6\: eV \) to \( 4.6\: eV \))
and symmetries (including \( D_{2d} \), \( C_{3v} \), \( C_{2v} \)
and \( T_{2d} \)), including some outwards relaxations (although
the majority favour inwards relaxation). In particular, Puska \emph{et
al}\cite{PuskaPPN98} thoroughly reviewed the previous theoretical
studies and also performed a sequence of \emph{ab initio} supercell
calculations using different system sizes and sampling techniques.
They found a large spread in possible answers, which they attributed
to the energy dispersion of the vacancy-induced deep levels, being
therefore particularly sensitive to details of the Brillouin zone
sampling scheme used.

Three key quantities of interest are:

\begin{enumerate}
\item the vacancy formation energy, which for a neutral vacancy in a supercell
is defined as\begin{equation}
\label{eqn: neutral vacancy formation energy}
E_{V}=E_{N-1}-\left( \frac{N-1}{N}\right) E_{N}
\end{equation}
where \( E_{N} \) is the total energy of the defect-free \( N \)
atom supercell, etc., 
\item the symmetry of the defect, and
\item the volume of the defect (actually the tetrahedron formed by the positions
of the four atoms \( \mathbf{r}_{1}\ldots \mathbf{r}_{4} \) surrounding
the vacancy)
\end{enumerate}
\begin{equation}
\label{eqn: volume of vacancy}
V=\frac{1}{6}\left| \left( \mathbf{r}_{4}-\mathbf{r}_{1}\right) \bullet \left( \mathbf{r}_{2}-\mathbf{r}_{1}\right) \times \left( \mathbf{r}_{3}-\mathbf{r}_{1}\right) \right| 
\end{equation}

\section{Method}

\label{sec: method}As an illustration of the methodology, we perform
what we believe to be the best converged \emph{ab initio} calculation
of the neutral silicon vacancy yet undertaken. As discussed in section
\ref{sec:review}, this is not the first time such a calculation has
been attempted. However, there has been a lot of scatter in the theoretical
calculations, even within the same paper in some instances! We seek
to explain the origin of this scatter, and in so doing, produce a
definitive answer for the neutral silicon vacancy formation energy
and the structure of the lattice relaxation around the vacancy. Our
calculations are performed with the CASTEP\cite{MSI99} code using
the PW91\cite{Perdew91} generalized gradient approximation (GGA)
for the exchange-correlation functional (which has been shown in many
previous defect studies to result in very accurate structures) with
a plane-wave basis set. We use a Vanderbilt ultrasoft pseudopotential\cite{Vanderbilt90}
for silicon which has been widely used and tested previously, e.g.
it gives the cubic lattice constant as \( 5.440 \) \AA{ } which compares
very favourably to the experimental value of \( 5.429 \) \AA{ } \( (+0.2\%) \).
For simplicity we have therefore fixed the lattice constants at the
value of the experimental lattice constant in all calculations.

As a measure of the reliability of the \emph{ab initio} scheme used,
we repeated certain calculations using the same CASTEP code but with
three different exchange-correlation functionals - the PW91 GGA as
mentioned above, and also the Ceperley-Alder\cite{CeperleyA80} local
density approximation (LDA) and the PBE\cite{PerdewBE96} GGA. The
same calculations were also repeated with an older norm-conserving
pseudopotential\cite{BacheletHS82,Hamann89,KleinmanB82} for silicon
that has been part of the standard CASTEP distribution for many years
and has been widely tested.

In the following sections, we will describe our methodology and illustrate
it with the neutral silicon vacancy calculation for definiteness.
Most of what follows can be applied to any supercell calculation,
but where there are parts of the discussion that are specific to silicon,
these will be be clearly highlighted.

Note that it is an often overlooked fact that, as we shall be relaxing
the atoms around the defect using forces derived from an \emph{ab
initio} calculation, we must ensure that the \emph{ab initio} calculation
is fully converged \emph{before} we start to consider any atomic relaxation.
That is, we must separately converge the electronic structure at fixed
atomic positions, before we can have any confidence in the forces
on the atoms being correct. Only then is it appropriate to attempt
to converge the atomic relaxation around the defect.

\subsection{Basis Set Size Convergence}

It is well known that the variational principle ensures that the total
ground state energy of a system will monotonically decrease as the
size of the basis set is increased. With some basis sets, it is difficult
to systematically improve the quality of the basis set - however,
with a plane-wave basis set as used here (which is often used with
supercell calculations) this is not a problem. We can associate an
energy with each plane-wave basis function (\( \left| \phi _{\mathbf{g}}\right\rangle \sim e^{i\mathbf{g}.\mathbf{r}} \)
where \textbf{g} is a reciprocal-space lattice vector), and so by
using all possible basis functions up to some maximum energy \( E_{cut} \)
we may characterize the size of the plane-wave basis set used. Therefore,
the basis set may be systematically improved by simply increasing
\( E_{cut} \) with a corresponding decrease in the total energy of
the system. It is a feature of plane-wave basis sets that typically
very large basis set sizes are required to achieve a reasonable tolerance
for the convergence of the total energy of a system, e.g. \( 1\: meV/atom \).
Therefore the pseudopotential approximation\cite{BacheletHS82,Hamann89,KleinmanB82,Vanderbilt90}
is invariably used which enable us to reduce the number of electrons
in the problem, and also to reduce the size of the basis set used
without effecting the accuracy of the treatment of the electrons outside
the core of the atoms. 

It is also well known that whilst the total energy of a given system
might converge slowly with increasing basis set size, the energy difference
between two similar systems is much more rapidly convergent due to
the cancellation of systematic errors. This is often used to discriminate
between two competing atomic arrangements with the same atoms in the
same supercell. However, in this work, we shall consider not the energy
difference, but the defect formation energy. Note that the variational
principle does not apply to such formation energies, and so we are
no longer guaranteed monotonic convergence, although this is often
seen in practice. We therefore start the calculation by converging
the unrelaxed defect formation energy as a function of basis set size,
for a reasonably small system. Everything else is kept fixed, e.g.
supercell size, sampling of reciprocal space for the Brillouin zone
integration, pseudopotentials, etc. In the case of the silicon vacancy,
we use the vacancy formation energy as defined in equation \ref{eqn: neutral vacancy formation energy}.
This necessitates calculating the total energy of the vacancy free
system (\( N \) atoms) and the vacancy system (\( N-1 \) atoms)
which we shall perform with all atoms fixed at the perfect lattice
coordinates. We shall then use the same cutoff energy (basis set size)
for all subsequent calculations unless otherwise noted.

For the neutral silicon vacancy, we compare the 16 silicon atom supercell
with the 15 atom supercell with vacancy. All atoms are kept at the
crystal positions with no relaxation. The Brillouin zone integration
is performed using a 2x2x2 Monkhorst-Pack grid. For this initial part
of the calculation, we work with a vacancy formation energy convergence
tolerance of \( 0.01\: eV \) and it is readily shown that with the
ultrasoft pseudopotential used that this corresponds to \( E_{cut}\sim 120\: eV \).

\subsection{Brillouin Zone Integration Convergence}

We perform the Brillouin zone integration using the method of special
\textbf{k}-points. Due to the localized nature of a defect in a (potentially
large) supercell, it is important to have a fully converged integration
here. This is the basis of the explanation for the difficulty of the
calculation as given by Puska \emph{et al}.\cite{PuskaPPN98} If we
use the simplest Monkhorst-Pack sampling scheme,\cite{MonkhorstP76}

\begin{equation}
\label{eqn: simple MP sample points}
\mathbf{k}=\frac{2\pi }{a}\left( \frac{q_{x}}{2q^{max}_{x}},\frac{q_{y}}{2q^{max}_{y}},\frac{q_{z}}{2q^{max}_{z}}\right) 
\end{equation}
with\[
q_{x}=\left\{ \begin{array}{cc}
\; \; \; 0,\pm 2\ldots \pm \left( q_{x}^{max}-1\right)  & q_{x}^{max}\, odd\\
\pm 1,\pm 3\ldots \pm \left( q_{x}^{max}-1\right)  & q_{x}^{max}\, even
\end{array}\right. \]
 etc., then we may easily converge the defect formation energy, at
fixed basis set size and system size, as a function of the density
of sample points in reciprocal space.

In order to maximize the separation of the defect from its periodic
image, we choose supercells that have the same nearest neighbour defect-defect
distances and sample uniformly in each direction in reciprocal space,
and so \( q^{max}_{x}=q^{max}_{y}=q^{max}_{z}=q \). We may therefore
systematically improve the convergence of the Brillouin zone integration
by simply increasing \( q \). The basic number of special points
in the grid is then \( q^{3} \). In order to minimize the number
of special points at a given value of \( q \) we apply the symmetry
operations of the supercell (\emph{not} the point-group symmetry of
the defect-free crystal lattice), and therefore work with a weighted
set of symmetrized points. The reduction in the number of points depends
on the symmetry of the supercell and the value of \( q \).

However, this will in general lead to very slow convergence, with
marked oscillations in both the total energy and the defect energy
as \( q \) is increased. This gives rise to the popular belief that
{}``odd \( q \) grids are less efficient than the corresponding
even \( q+1 \) grid''. However, this is a failure of the implementation
of the grid, not the general Monkhorst-Pack method. This can be seen
by calculating the radius of exact integration in reciprocal space
for different values of \( q \) for some of the most common supercells
- simple cubic (SC), body-centred cubic (BCC) and face-centred cubic
(FCC) as seen in table \ref{table: simple supercells, no offsets}.
For an ideally efficient sampling scheme, the integration should be
exact out to a radius given by \( R=q \), but it can be seen from
table \ref{table: simple supercells, no offsets} that this is only
achieved for all \( q \) for the simple cubic supercell, and for
even-valued \( q \) for the face-centred cubic supercell.

This flaw was overcome in the basic Monkhorst-Pack scheme by the possible
inclusion of a rigid offset, \( \mathbf{k}_{0} \), of the sampling
grid from the origin of reciprocal space. This offset is often ignored,
but is essential to achieve the full efficiency of the scheme. The
use of the optimal offset for a given supercell symmetry and value
of \( q \) removes the oscillations in the total energy and consequently
accelerates the convergence of the Brillouin zone integration. A comprehensive
set of these optimal offsets were derived by Moreno and Soler\cite{MorenoS92}
(and independently by MIJP) but the significance of their work has
not been widely appreciated. For completeness, the optimal offsets
for the supercells considered in this work are shown in table \ref{table: simple supercells, with offsets}.

We find that the use of a Monkhorst-Pack grid with the optimal offset
is more widely applicable and technically superior to other schemes
proposed, such as ref. \onlinecite{MakovSP96}. Using this approach
it is now possible to approach convergence of the Brillouin zone integration
in a consistent manner for any value of \( q \) required. Note, that
for certain supercells, such as body-centred cubic, the use of offsets
is beneficial for all values of \( q \), whereas for others, such
as face-centred cubic, it is only beneficial for odd-valued \( q \).

When comparing the quality of the Brillouin zone integration for two
different sized or different shaped systems, it is not the value of
\( q \) or the number of special points that should be compared,
but rather the density of symmetry-unfolded special points in reciprocal
space. Note that this comparison will be simplest for two different
systems if in each case the sampling scheme is equally efficient (e.g.
\( R=q \)). Therefore we recommend the use of offsets at all stages
in this methodology when calculating the convergence of the electronic
structure.

\begin{figure}
{\centering \resizebox*{0.9\columnwidth}{!}{\rotatebox{270}{\includegraphics{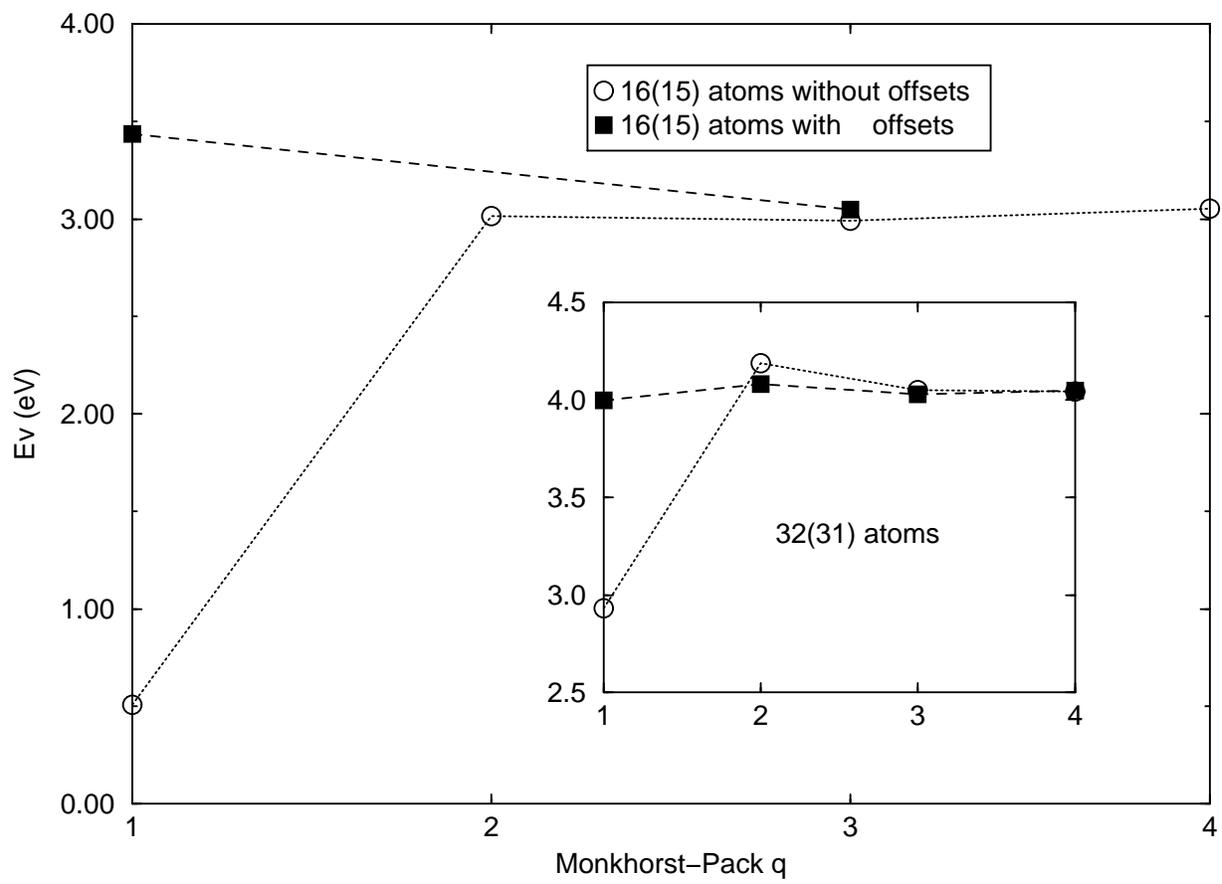}}} \par}

\caption{\label{fig: si16_BZ_convergence}Convergence of unrelaxed vacancy
formation energy w.r.t Brillouin zone sampling, for 16/15 atom system
at \protect\( E_{cut}=120\: eV\protect \). The inset shows the corresponding
convergence for the 32/31 atom system.}
\end{figure}

Therefore, for the neutral silicon vacancy, we converge the Brillouin
zone sampling using the 16/15 atom FCC supercells and \( E_{cut}=120\: eV \).
As shown in table \ref{table: simple supercells, with offsets}, there
is no advantage to using offsets with even-\( q \) grids and FCC
supercells but there is a difference for odd-\( q \) grids. We therefore
perform the odd-\( q \) grid calculations twice, once with and once
without offsets, and the results of increasing \( q \) on the vacancy
formation energy, both with and without offsets, are shown in figure
\ref{fig: si16_BZ_convergence}. There is clearly a dramatic improvement
for \( q=1 \) where using an offset shifts the sampling away from
the gamma point, but it may not appear too dramatic for other values
of \( q \) (although it may be hard to see from the figure, in fact
the convergence of \( q=3 \) w.r.t. \( q=4 \) is significantly improved
from \( -0.061\: eV \) to \( -0.005\: eV \)). Moreover, table \ref{table: simple supercells, with offsets}
suggests that the benefits of using offsets will be most marked with
BCC supercells. Therefore, for illustration, we also repeat the Brillouin
zone convergence calculation with the 32/31 atom BCC supercells and
\( E_{cut}=120\: eV \), again both with and without offsets. The
results are shown as the insert to figure \ref{fig: si16_BZ_convergence},
and show a marked improvement in convergence with offsets. If we again
apply a vacancy formation energy convergence tolerance of \( 0.01\: eV \)
it can be seen that this corresponds to a Brillouin zone sampling
density of \( \leq 0.033 \) \AA\( ^{-1} \). This was therefore used
as our sampling density in all subsequent calculations.

\subsection{Supercell Finite Size Convergence}

Having fully converged the electronic structure calculation for the
unrelaxed defect in a given size system, we can proceed to converge
the effects of the finite size supercell. This is the key difference
between the supercell and the cluster approaches. With a cluster,
we need to minimize the interaction between the defect and the surface
of the cluster, but here, with a supercell, we need to minimize the
interaction between the defect and its own periodic images. Hence
the requirement to converge the supercell size. For insufficiently
large supercells, there will be an appreciable overlap between the
defect and its own images, resulting in an error in the overall charge
density of the system, and hence the total energy and the forces on
the atoms. The obvious solution to this is to repeat the defect formation
energy calculation in different sized supercells, using an equivalent
sized basis set (e.g. same plane-wave cutoff energy) and same Brillouin
zone sampling density.

\begin{figure}
{\centering \resizebox*{0.5\columnwidth}{!}{\rotatebox{270}{\includegraphics{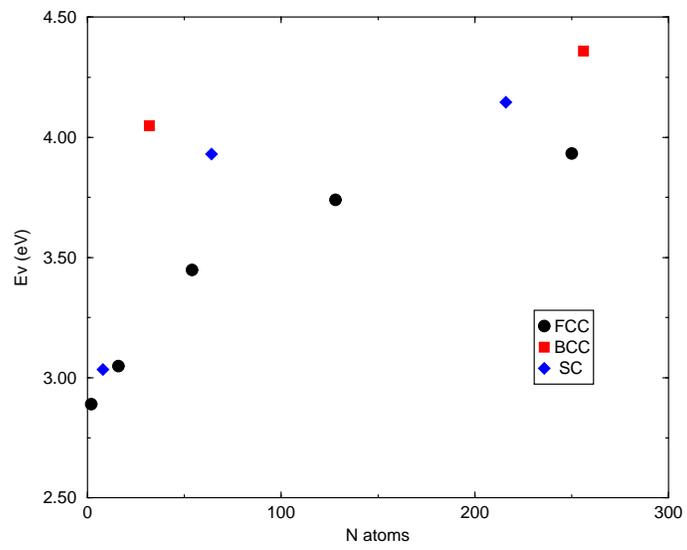}}} \par}

(a)

{\centering \resizebox*{0.5\columnwidth}{!}{\rotatebox{270}{\includegraphics{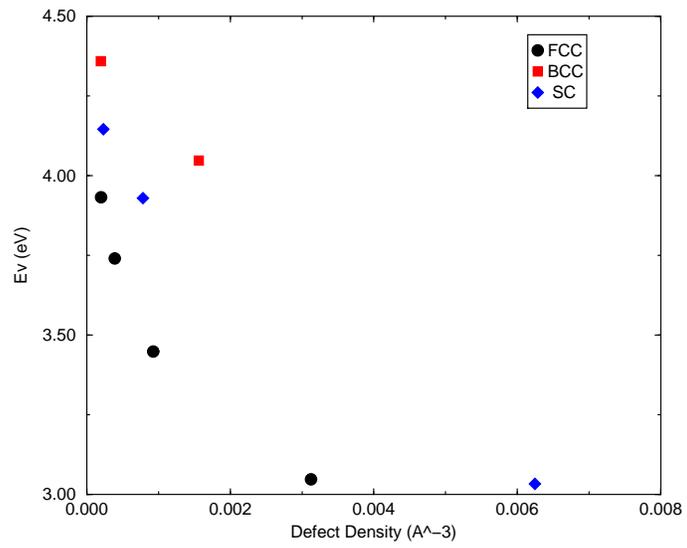}}} \par}

(b)

\caption{\label{fig: si_vacancy_N}Variation of unrelaxed defect formation
energy at constant basis set size and Brillouin zone sampling. (a)
shows the variation with system size, and (b) shows the variation
with density. The different symmetry supercells are clearly separated
in (b).}
\end{figure}

For the neutral silicon vacancy therefore, we considered all possible
SC, BCC and FCC supercells with between 2 and 256 atoms in the vacancy
free system. The actual value of \( q \) used and the corresponding
sampling density are summarized in table \ref{table: supercell list}.
The unrelaxed vacancy formation energy, at full Brillouin zone sampling
convergence, for each different supercell is plotted in figure \ref{fig: si_vacancy_N}(a)
as a function of the number of atoms in the corresponding vacancy
free system. This is a common way of presenting such information,
yet this figure appears confusing, with no obvious trend apparent
in the convergence of the vacancy formation energy with system size.
However, separating the different points according to the supercell
symmetry suggests that there may be a trend, but that this is not
the best way to present such data. This is because if we simply order
the different possible supercells in terms of the total number of
atoms (or equivalently, the defect-image distance), we will be misled
as the \emph{defect density} will be changing in a non-monotonic manner.
Instead, figure \ref{fig: si_vacancy_N}(b) plots the vacancy formation
energy against the defect density, which clearly separates out the
different supercell symmetries. This now eliminates the apparent scatter
in figure \ref{fig: si_vacancy_N}(a) and instead three clear monotonic
trends are evident, one for each supercell symmetry. These trends
all appear to converge to the same value, \( \simeq 4.40\: eV \),
in the limit of infinite supercell size (defect density = 0) as would
be expected. This therefore explains a common source of the scatter
seen between and within the different theoretical studies of the silicon
vacancy to date. This effect will obviously also apply to any other
supercell defect study.

What then is the origin of the different rates of convergence of the
defect formation energy for different symmetry supercells? A simple
tight-binding model of nearest-neighbour interactions (with hopping
matrix element = \( \gamma \left( a\right)  \) where \( a \) is
the separation of nearest-neighbours) is given in many standard texts,
e.g. ref. \onlinecite{AshcroftM76_10}. In this generic model, a band
will be formed with a characteristic bandwidth of \( 12\gamma  \)
for SC supercells, and \( 16\gamma  \) for BCC or FCC supercells
with the same defect separation. This can be attributed to the effects
of geometry as well as the different number of nearest neighbours
in the different supercells. It might therefore be expected that SC
supercells were to be preferred in general for defect calculations
as they have the least defect-defect interaction (smallest bandwidth)
at a given defect separation.

Indeed, some evidence for this is seen in figure \ref{fig: si_vacancy_N}(b),
where it can be seen that SC supercells are converging at a faster
rate than FCC ones. Indeed, it appears that the 64/63 atom SC supercell
gives a comparable representation of an isolated unrelaxed vacancy
to the 250/249 atom FCC supercell, which can be attributed in part
to the number of nearest neighbour defects. However, this model does
not explain why (for the neutral silicon vacancy) the 32/31 atom BCC
supercell gives an even better representation (i.e., the energy is
closer to the zero density limit) than either the 64/63 atom SC supercell
or the 250/249 atom FCC supercell which implies that this simple tight-binding
model is of limited usefulness.

There must therefore be a further consequence of the supercell symmetry
that has not been considered so far. This is that the defect-defect
interaction will have the directionality of the supercell which may
or may not be commensurate with the underlying crystal symmetry. This
causes a perturbation in the electronic charge density that is another
finite-size effect which must vanish in the limit of a sufficiently
large supercell. A rigorous analysis of this would involve calculating
the density-density response function (see e.g. ref. \onlinecite{Springborg00}).
However, the effect becomes apparent if we simply plot the charge
density difference between the unrelaxed vacancy and vacancy-free
systems.

\begin{figure*}
{\centering \resizebox*{0.3\textwidth}{!}{\includegraphics{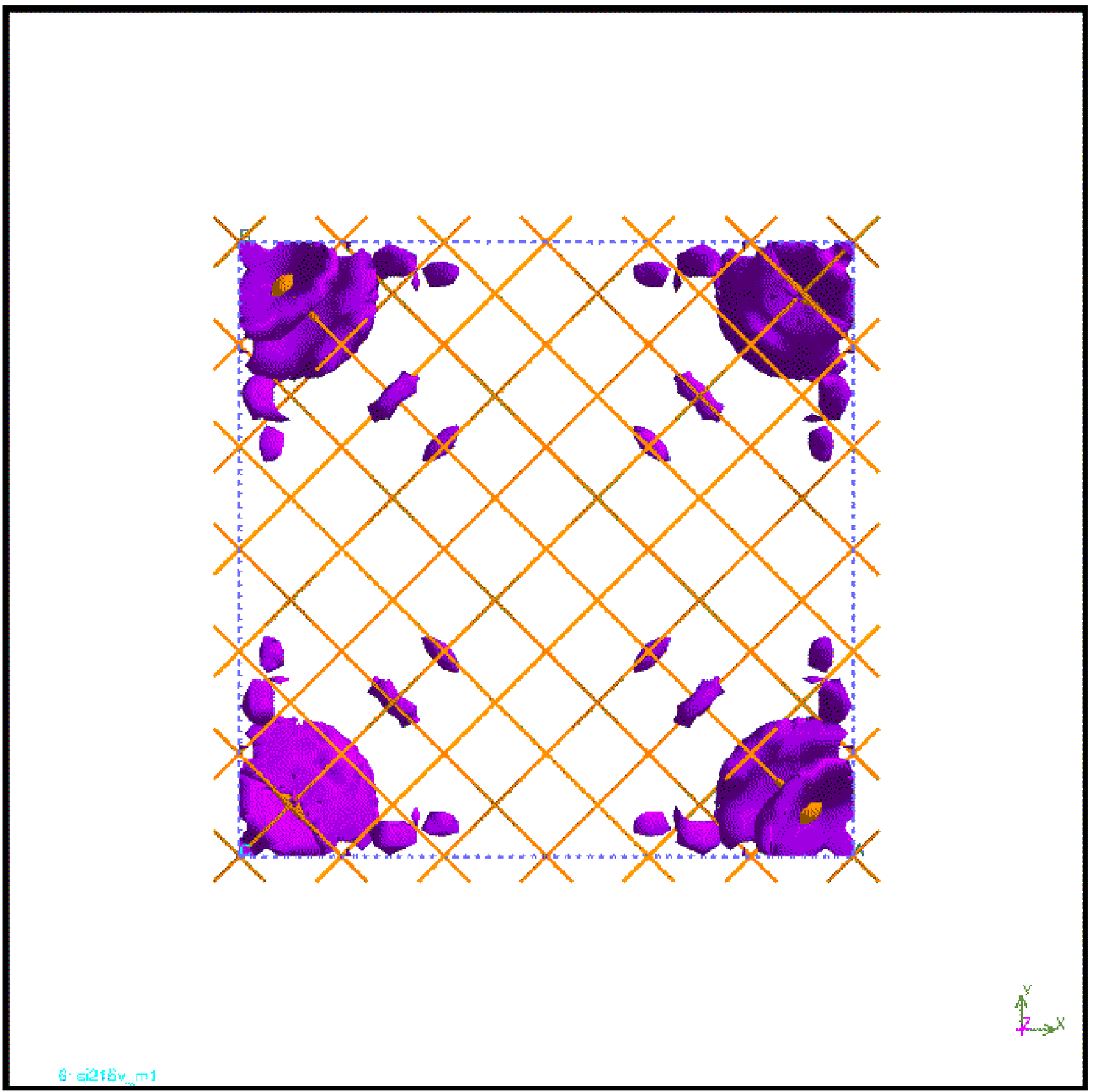}} \resizebox*{0.3\textwidth}{!}{\includegraphics{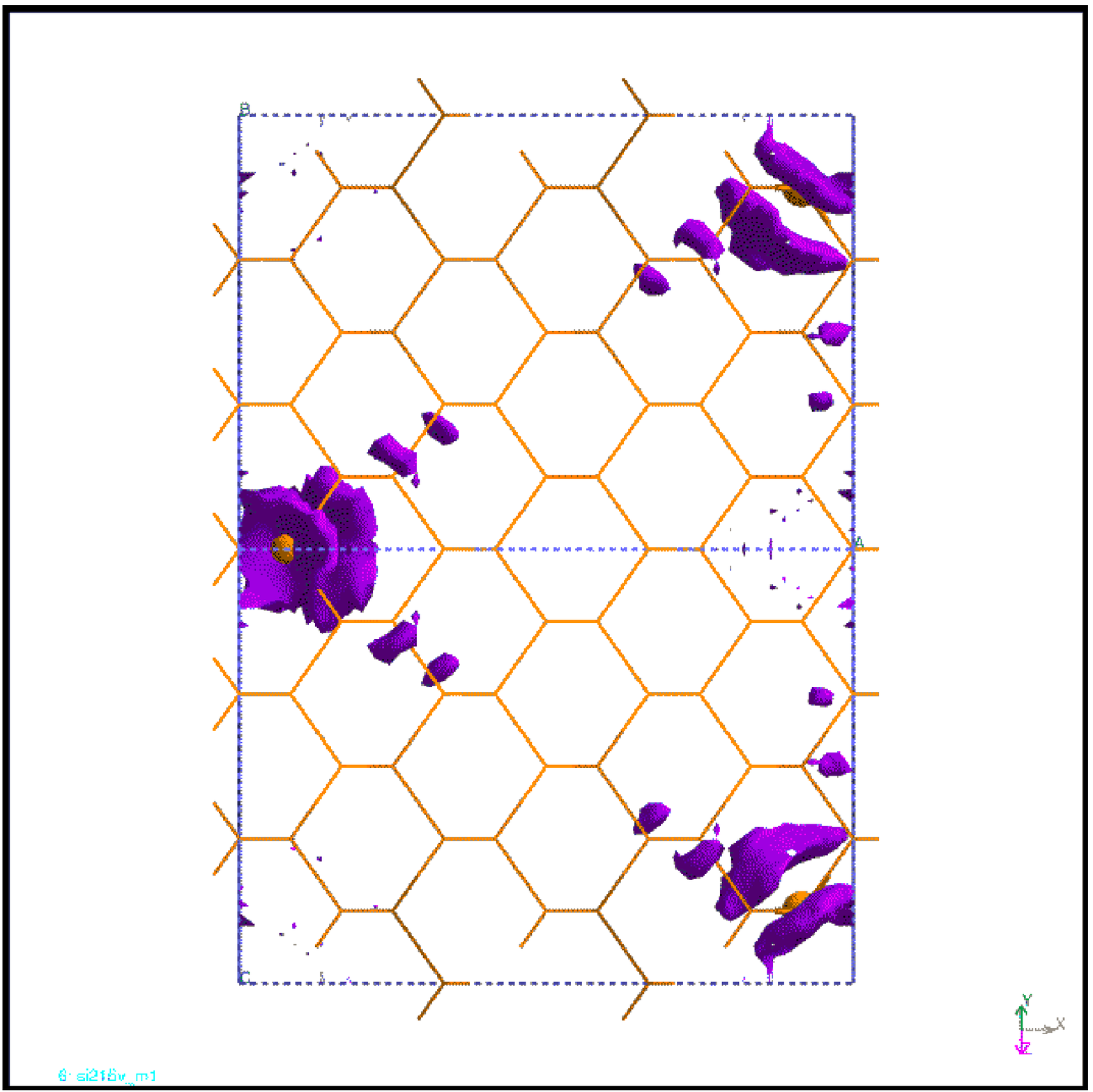}} \resizebox*{0.3\textwidth}{!}{\includegraphics{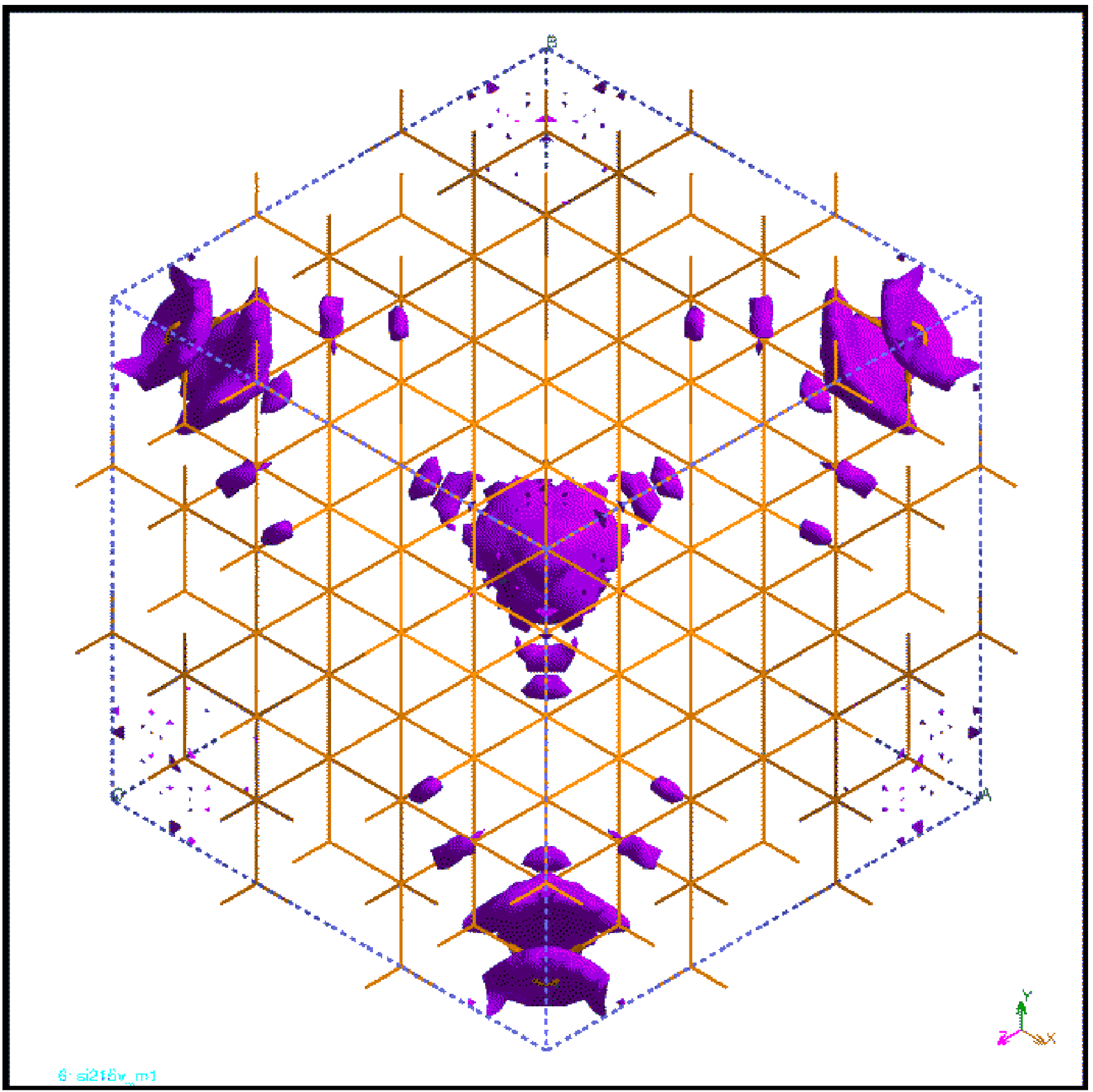}} \par}

\caption{\label{fig: si215v_p002}Charge density difference iso-surface at
\protect\( \rho =0.002\: eV/\protect \)\AA \protect\( ^{3}\protect \)
between the unrelaxed 216 and 215 atom SC supercells. Leftmost figure
is viewed along the <001> direction, central figure is along the <011>
direction, and rightmost figure is along the <111> direction.}
\end{figure*}

\begin{figure*}
{\centering \resizebox*{0.3\textwidth}{!}{\includegraphics{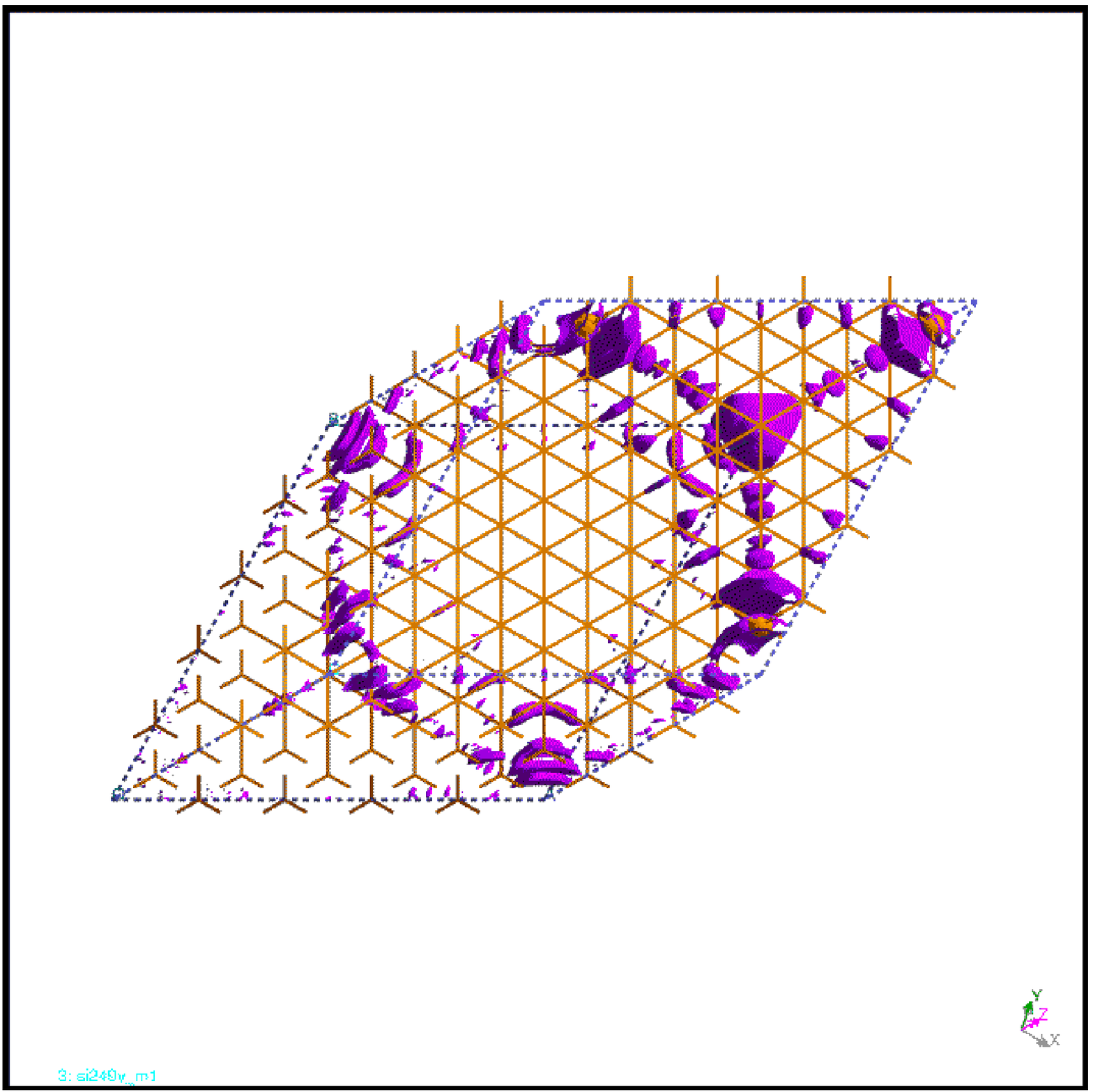}} \resizebox*{0.3\textwidth}{!}{\includegraphics{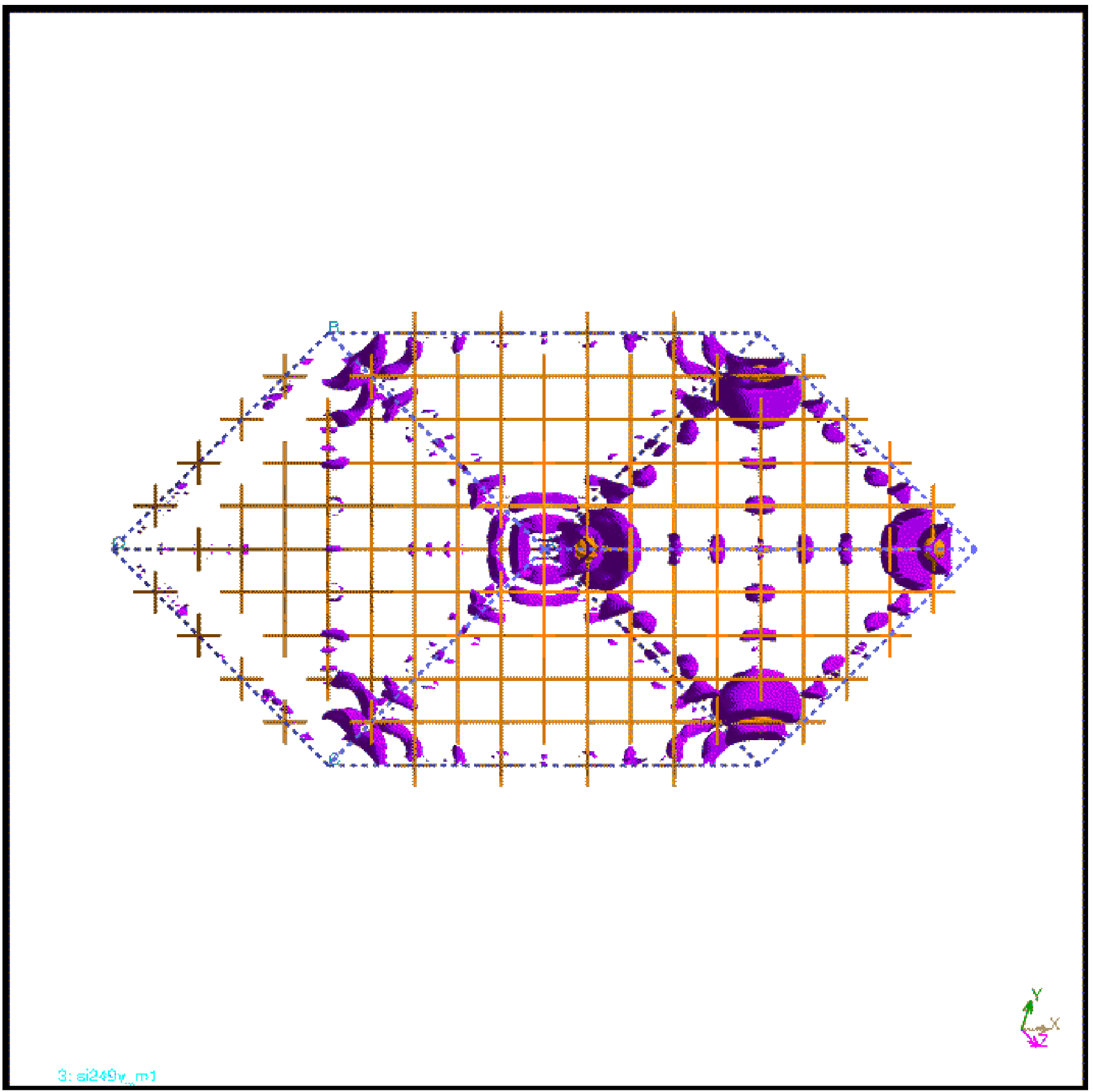}} \resizebox*{0.3\textwidth}{!}{\includegraphics{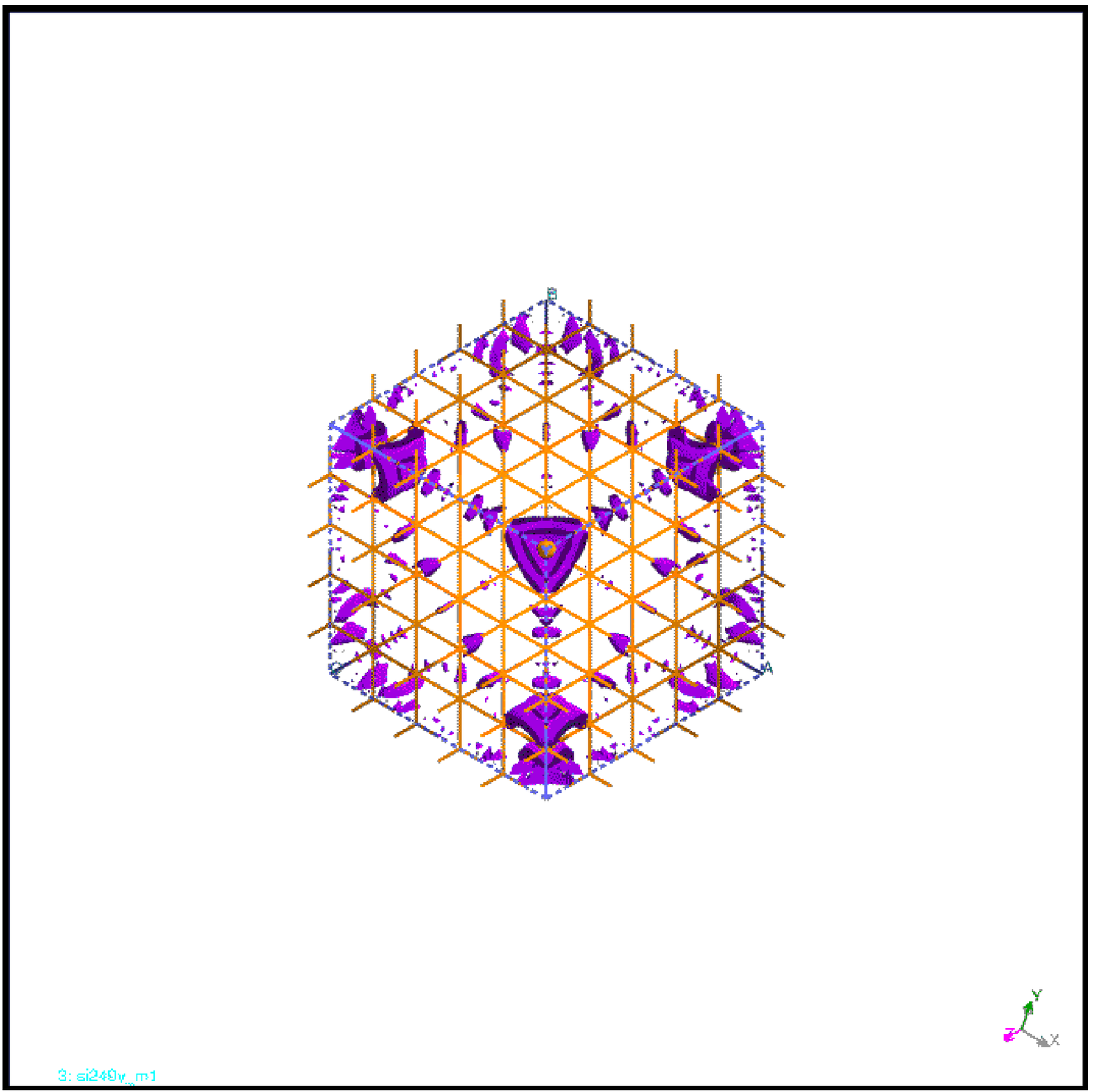}} \par}

\caption{\label{fig: si249v_p002}Charge density difference iso-surface at
\protect\( \rho =0.002\: eV/\protect \)\AA \protect\( ^{3}\protect \)
between the unrelaxed 250 and 249 atom FCC supercells. Leftmost figure
is viewed along the <001> direction, central figure is along the <011>
direction, and rightmost figure is along the <111> direction.}
\end{figure*}

\begin{figure*}
{\centering \resizebox*{0.3\textwidth}{!}{\includegraphics{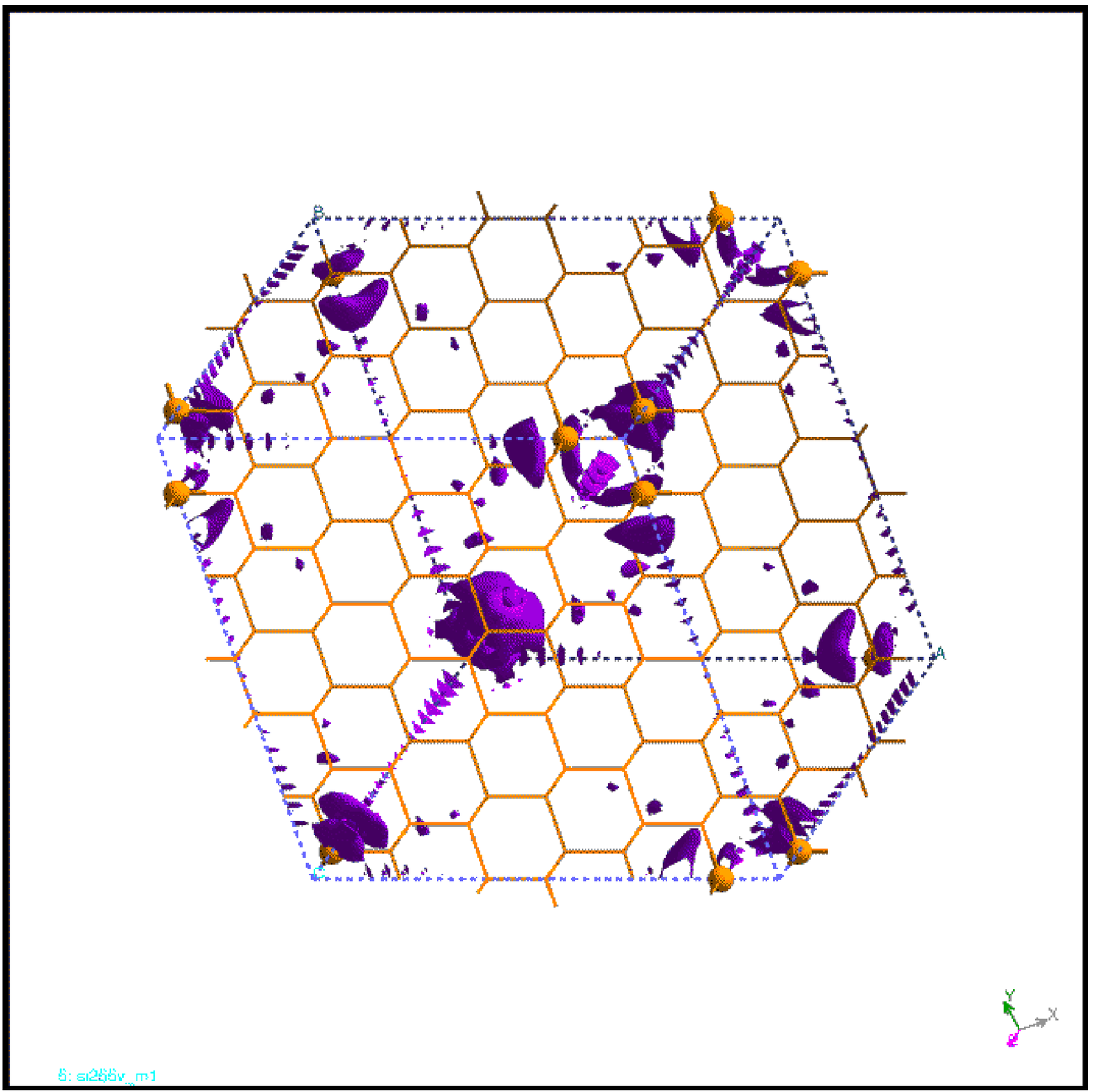}} \resizebox*{0.3\textwidth}{!}{\includegraphics{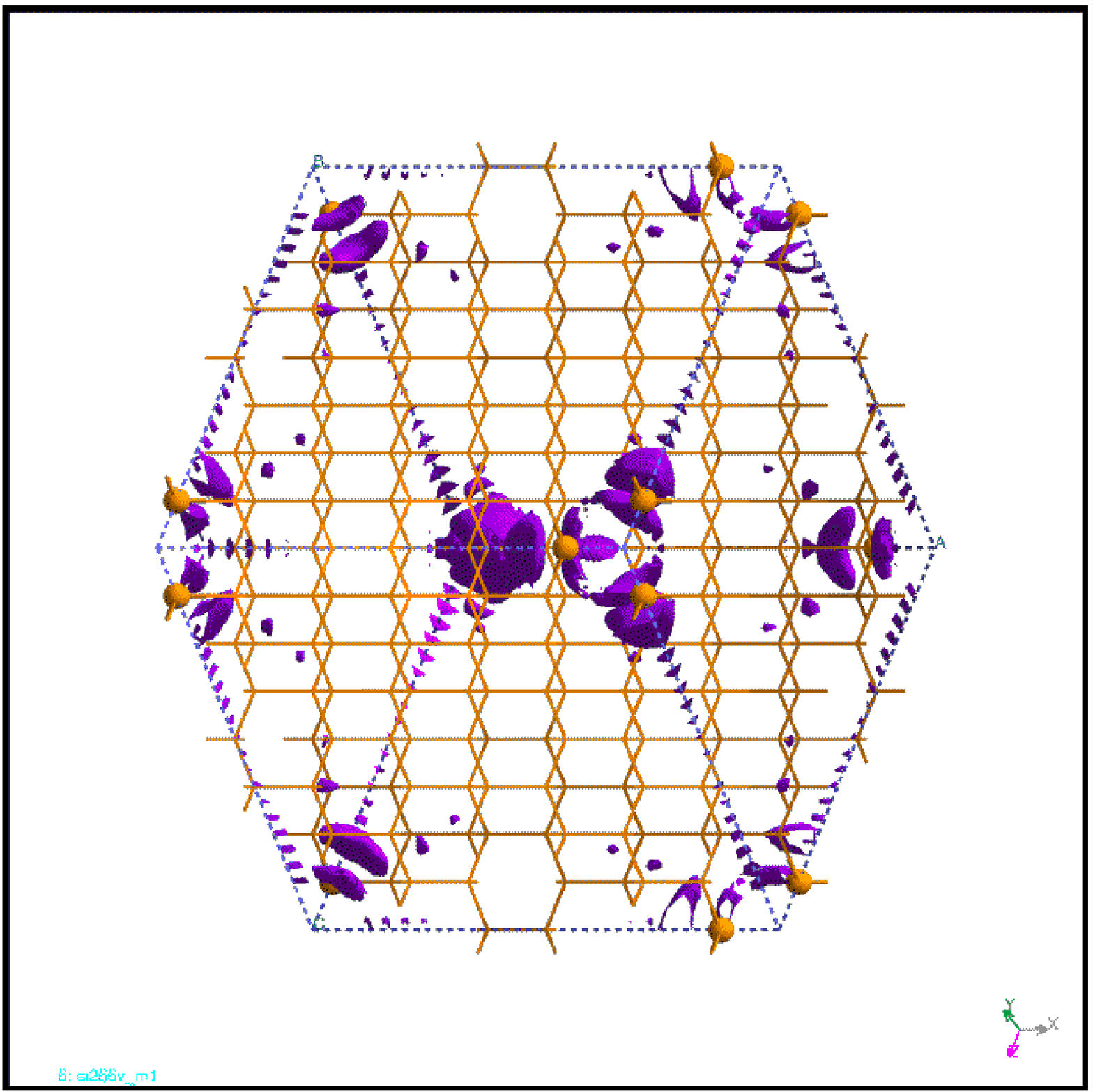}} \resizebox*{0.3\textwidth}{!}{\includegraphics{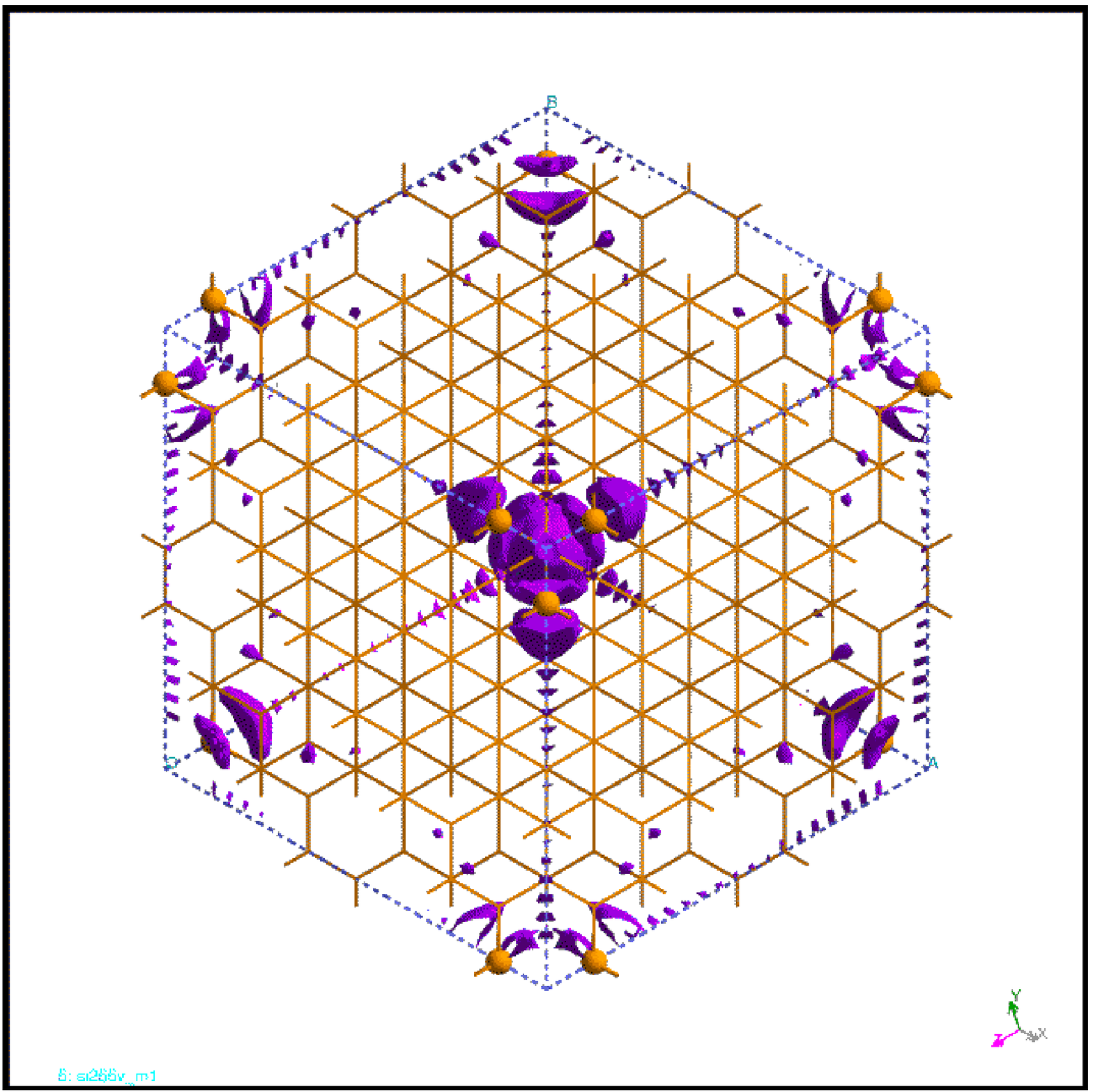}} \par}

\caption{\label{fig: si255v_p002}Charge density difference iso-surface at
\protect\( \rho =0.002\: eV/\protect \)\AA \protect\( ^{3}\protect \)
between the unrelaxed 256 and 255 atom BCC supercells. Leftmost figure
is viewed along the <001> direction, central figure is along the <011>
direction, and rightmost figure is along the <111> direction.}
\end{figure*}

For the case of the neutral silicon vacancy, such plots are shown
for the 216/215 atom SC supercell in figure \ref{fig: si215v_p002},
the 250/249 atom FCC supercell in figure \ref{fig: si249v_p002} and
the 256/255 atom BCC supercells in figure \ref{fig: si255v_p002}.
In the SC supercell we see that there is a localized charge density
difference around the vacancy, and then a longer ranged component
which spans the supercell which is clearly aligned with the <100>
directions. Similarly, in the FCC supercell the long ranged component
is along the <110> directions and in the BCC supercell it is along
the <111> directions. The normal silicon-silicon bonds are in <111>
directions which then explains why the BCC supercell is superior for
silicon defects - the spurious charge movements caused by the finite
supercell size effect are commensurate with the underlying charge
density of the system and hence make little difference to the total
energy. This is not the case in the FCC and SC supercells where it
can be seen that there have been spurious charge movements in the
interstitial regions where the charge density is naturally lower,
which therefore has a more significant effect. This clearly shows
that it is not sufficient to simply increase the size of the system
to get a {}``better'' answer, which contributes to the confusion
in some earlier studies of the silicon vacancy. The directionality
effect of the supercell symmetry will also apply in general to any
other defect system, although the detailed considerations will, of
course, vary.

Unfortunately, there are only two BCC supercells in the range 2-256
atoms (32 and 256 as in table \ref{table: supercell list}), which
would therefore seem to limit our ability to make judgments about
the efficacy of BCC supercells for silicon defects. As a further test,
the calculation of the unrelaxed neutral silicon vacancy was then
repeated for the next BCC supercell, which corresponds to 864 atoms
with a defect density of \( 0.000058 \) \AA\( ^{-3} \). Again, the
same basis set cutoff, Brillouin zone sampling density and offset
were used. The corresponding unrelaxed defect formation energy was
\( 4.358\: eV \). This confirmed the prediction about the infinite
supercell size limit, and shows that the 256 atom BCC supercell has
converged (to better than \( 0.002\: eV \)) the electronic structure
of the unrelaxed neutral silicon vacancy w.r.t. finite supercell size.

In the case of charged defects, the effects of the finite size supercell
will be even more marked due to the long-ranged nature of the Coulomb
interaction. Specialised energy correction schemes have been introduced
(e.g. \emph{}Makov-Payne\cite{MakovP95}) that accelerate the convergence
of the total energy with increasing supercell size.

\subsection{Hellman-Feynman Forces Convergence}

Having finally fully converged all the necessary factors in the unrelaxed
defect formation energy, we can now be confident that we have an accurate
representation of the ground state electronic wavefunction. We may
now use the Hellman-Feynman theorem to calculate the forces on the
atoms and hence start to relax the defect. However, it must be noted
that we converged the basis-set size using an energy difference calculation.
The variational principle assures us that the ground state energy
is correct to second-order errors in the ground state wavefunction,
but the forces will only be correct to first-order errors. Also, as
noted previously, an energy difference will converge more rapidly
than the total energy.

The advantage of using the defect formation energy criterion in the
early stages of this methodology is that it produces a smaller basis
set which makes the other (unrelaxed structure) convergence calculations
rapid. This can produce significant savings, as to be sure of convergence
it is often necessary to go to one size of calculation beyond that
at which convergence first appears, as for example when converging
the finite size supercell effect in the neutral silicon vacancy where
an 864 atom BCC supercell was evaluated.

In order to produce accurate forces therefore, it is necessary to
converge the basis set size with respect to the forces and so we choose
the RMS force of the unrelaxed defect structure as a simple scalar
parameter to converge. This additional convergence is especially important
for defect calculations, as it is often found that the energy surface
around a defect is very flat, and so particularly prone to errors
in the forces due to the use of under-converged basis sets. This sort
of effect can be easily detected by monitoring the direction of the
forces on each atom surrounding the defect as the basis-set size is
increased. Any tendency for this direction to change significantly
is a clear warning that there are serious systematic errors in the
forces due to basis-set incompleteness.

\begin{figure}
{\centering \resizebox*{0.9\columnwidth}{!}{\rotatebox{270}{\includegraphics{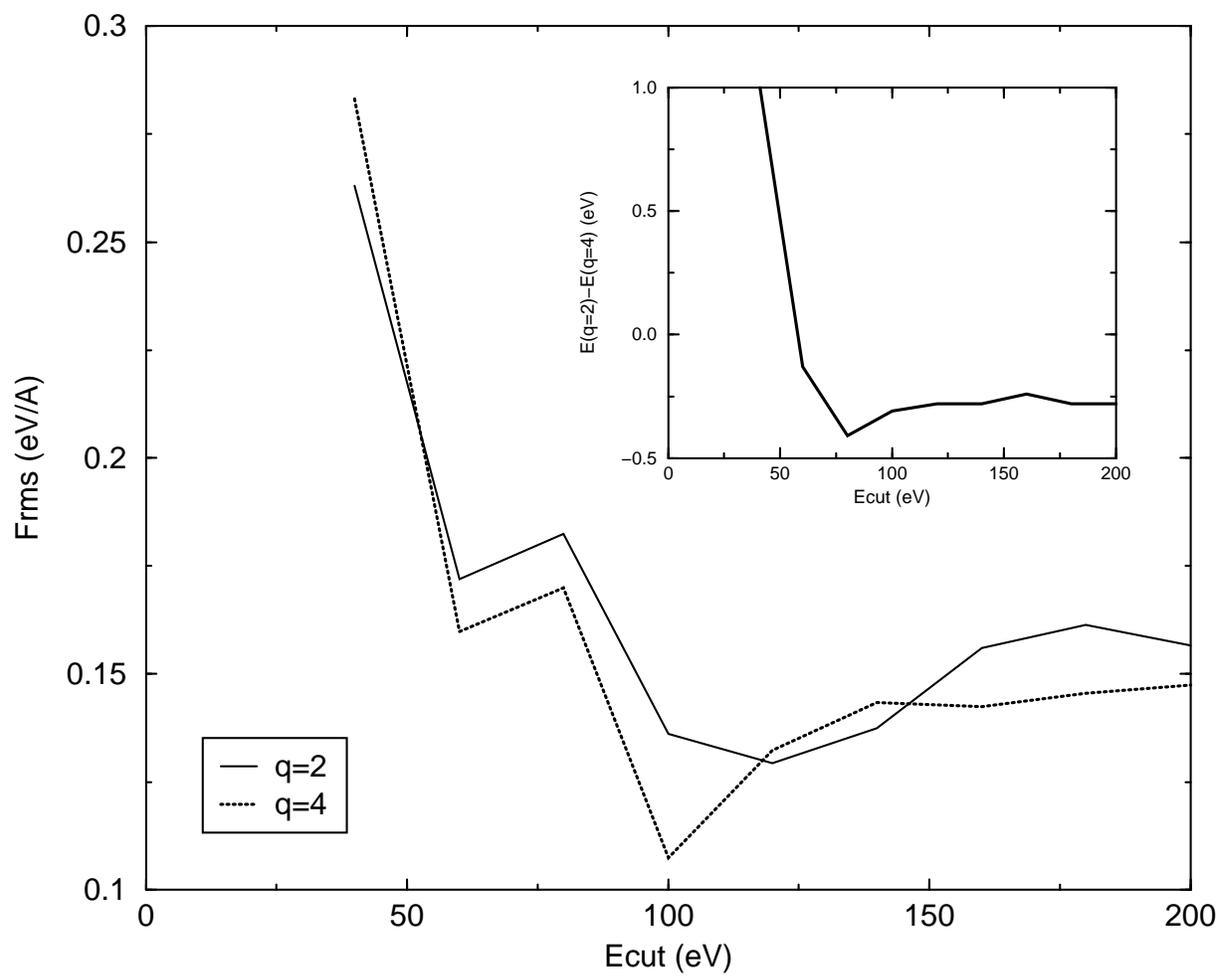}}} \par}

\caption{\label{fig: si31v_Frms}Variation of RMS force in 31 atom silicon
cell w.r.t basis set size at two different Brillouin zone sampling
densities, corresponding to different values of the Monkhorst-Pack
grid parameter \protect\( q\protect \). The inset figure shows the
corresponding difference in the total energy for the two different
values of \protect\( q\protect \). This clearly shows that whilst
it might appear that the total energy is adequately converged at \protect\( E_{cut}=120\: eV\protect \)
and \protect\( q=2\protect \), this is not sufficient for the forces.
In all subsequent relaxation calculations, a Brillouin zone sampling
density equivalent to that corresponding to \protect\( q=4\protect \)
in this calculation, and \protect\( E_{cut}=160\: eV\protect \) was
used.}
\end{figure}

An example calculation for the case of the neutral silicon vacancy
in the 32/31 atom BCC supercell is shown in figure \ref{fig: si31v_Frms},
where it can be seen that whilst from an energy calculation it appears
that \( E_{cut}=120\: eV \) and \( q=2 \) is reasonably converged,
this is not sufficient for the forces. Applying a criterion that the
RMS force must be converged to \( 0.005\: eV/ \)\AA{} (which is often
used as the convergence tolerance in high quality \emph{ab initio}
structural relaxations) we see that \( E_{cut}=160\: eV \) must be
used, and that a Brillouin zone sampling of \( q=4 \) (corresponding
to a density of \( 0.033 \) \AA\( ^{-1} \)) must be used.

\subsection{Atomic Relaxation Convergence}

Finally, we are now ready to relax the atomic structure around the
defect, using the forces derived from the systematically converged
\emph{ab initio} calculation. We move the atoms according to some
minimization algorithm, and stop when we have simultaneous satisfied
the various relaxation criteria to prescribed tolerances: e.g. convergence
of the total energy, the RMS force, and the RMS displacement of the
atoms between successive iterations. If we had been simultaneously
optimising the lattice parameters using \emph{ab initio} stresses,
then it would be appropriate to also check for convergence of the
stress on the supercell.

Note that we are often starting the atomic relaxation from a state
of relatively high symmetry. It may therefore be necessary to perturb
each atom by a small amount from the symmetry sites at the start of
the calculation in order to ensure symmetry breaking is possible in
the relaxation process. Also, because of the possibility of local
minima in the structure minimization, the calculation should be restarted
several times from different initial arrangements of atoms around
the defect (e.g. random symmetry breaking displacements, directed
relaxation inwards, directed relaxation outwards, etc) in order to
be sure that the minimized structure found is indeed the global minimum.

Note that proving that any particular minimum found is indeed the
global minimum is a difficult matter. More advanced techniques, such
as simulated annealing\cite{KirkpatrickGV83} or \emph{ab initio}
molecular dynamics\cite{CarP85}, are better adapted to exploring
the energy surface but at much increased computational cost. In practice,
if several independent starting configurations all converge to the
same answer, then that is usually sufficient to have a reasonable
amount of confidence that the structure found is (a close approximation
to) the global minimum.

There is also a popular belief that it is more efficient to relax
a structure using a small basis set to get an approximate structure
and then to increase the size of the basis set until there is no further
change, than to use a sufficiently large basis set throughout. The
neutral vacancy in silicon is a counter-example to that belief. If
the defect structure is relaxed using too small a basis set (e.g.
\( E_{cut}=120\: eV \)), then the systematic errors in the forces
cause the vacancy to relax \emph{outwards}. This outwards relaxation
is remarkably robust w.r.t. different perturbations of the surrounding
atoms prior to starting the relaxation, including gross inwards and
outwards distortions, and the final state is also locally stable w.r.t.
subsequent increases in the basis set size proving that it is a local
minimum. However, if the vacancy is relaxed using a larger basis set
(\( E_{cut}\geq 160\: eV \)) at all times, then the resulting relaxation
is \emph{inwards} which illustrates the importance of monitoring the
direction of the forces on the unrelaxed atoms surrounding the defect
as the basis-set size is increased, as suggested above. This inwards
relaxation is also robust w.r.t. a range of different starting configurations,
and the final minimized structure is \emph{lower} in energy than the
outwards-relaxed structure at the same basis set size. 

This might appear confusing at first, as the local potential energy
surface around each atom should be quadratic as silicon at low temperatures
is a harmonic crystal to a good approximation - hence the equilibrium
geometry ought to be reasonably insensitive to the detail of the calculation.
However, this result implies that changing the basis-set size (i.e.
reducing the systematic errors in the forces) causes a significant
change in the gradient of the potential energy surface around the
unrelaxed defect, i.e. the forces as seen in figure \ref{fig: si31v_Frms}.
So it is actually the boundaries of the different basins of attraction
for the relaxation minimizer that are being moved. 

This therefore explains another common source of the scatter seen
between the different theoretical studies of the neutral silicon vacancy
to date, and shows that the only way to relax the defect reliably
is to use the larger basis-set size at all stages in the relaxation
to reduce the systematic errors in the forces, and to check that the
resulting configuration found is the global and not just a local minimum.

When comparing the energetics of two different structures, it should
be borne in mind that experiments are usually conducted at finite
temperature, whereas energy minimisation strategies usually correspond
to zero temperature. This means that a true comparison should be based
upon free energies and not just total energies. The entropy difference
or the free energy difference between the structures can be obtained
by various techniques, such as thermodynamic integration using constrained
molecular dynamics (e.g. see ref \onlinecite{FrenkelS96_7} for details).
Another complication that arises with finite temperature, is that
the presence of nearby local minima will produce significant temperature
dependencies in many physical properties, whereupon it then becomes
important to know the location of these other minima and the saddle-points
separating them from the global minimum.

\subsection{Defect Structure Convergence}

However, even at this stage, there is still one more convergence criterion
to meet. The atomic relaxation around the defect may be quite long
ranged, and the pattern of relaxations must be contained within the
supercell. That is, if we consider successive shells of atoms around
the defect (i.e. all those atoms at a common distance from the defect
in the unrelaxed structure), then there should be negligible relaxation
for atoms beyond a certain distance from the defect - and certainly
before the largest shell allowed by the periodic boundary conditions
(i.e. half the defect-image separation). 

One way to provide an upper-bound on the relaxation energy is to perform
the atomic relaxation calculation in stages - that is, in the first
calculation to only relax those atoms in the first shell around the
defect, and then in successive calculations to increase the number
of shells allowed to relax, up to the largest allowed shell. Each
successive calculation will then provide an improved estimate of the
relaxation energy, and allow a simple determination as to whether
or not the relaxation has been properly contained within the finite
size supercell. This approach is known as {}``relaxation under a
constant strain field'' and is useful for calculating an upper-bound
on the relaxation energy in a small system, but has the disadvantage
that it might result in the system being trained into a local minima
which is not the global minimum. Therefore, the best approach is to
calculate the relaxation energy without any constraints - in which
case, if the supercell is large enough, there will be negligible relaxation
of the largest allowed shells. It is standard practice for relaxation
calculations to be repeated for different random perturbations of
the atom coordinates, to ensure that the same minimum structure is
reached each time.

As a cross-check that the supercell is large enough, and that spurious
symmetry effects (e.g. force cancellation in certain directions) have
not caused a misleading conclusion, the strain on the supercell should
be evaluated, and the volume allowed to relax as appropriate. However,
in a fixed volume calculation, there will often be a uniform breathing-mode
expansion or contraction of the further-out shells as the underlying
lattice accommodates the local relaxation around the defect. This
effect will tend to increase the apparent size of the relaxation and
cause a volume relaxation that may not be warranted. Therefore, to
assess convergence of the defect structure, we consider the relative
displacement of successive shells of atoms between the relaxed and
unrelaxed defect systems in a fixed volume calculation (as in ref.
\onlinecite{WangCH91}) and check that this is converged (to some appropriate
tolerance) \emph{before} the largest allowed shell allowed by the
periodic boundary conditions (i.e. half the defect-image separation). 

If it is found that the relaxation is not contained within this largest
allowed shell, then the supercell must be increased in size and the
above procedure repeated until this is no longer the case. Only then
can it be claimed that the calculation is representative of an isolated
defect. Of course, this might result in supercell sizes that are impracticable
with current computer resources. It has long been recognised that
the best way to improve supercell calculations is to use a larger
supercell, and for a long time the largest supercell practical for
studying the neutral silicon vacancy was suspected to be too small.
However, one of the conclusions of this study is that the best way
to improve a calculation is not just to increase the supercell size,
but to do so in an appropriate manner bearing in mind the interaction
of the supercell symmetry with the defect. 

\begin{figure}
{\centering \resizebox*{0.9\columnwidth}{!}{\includegraphics{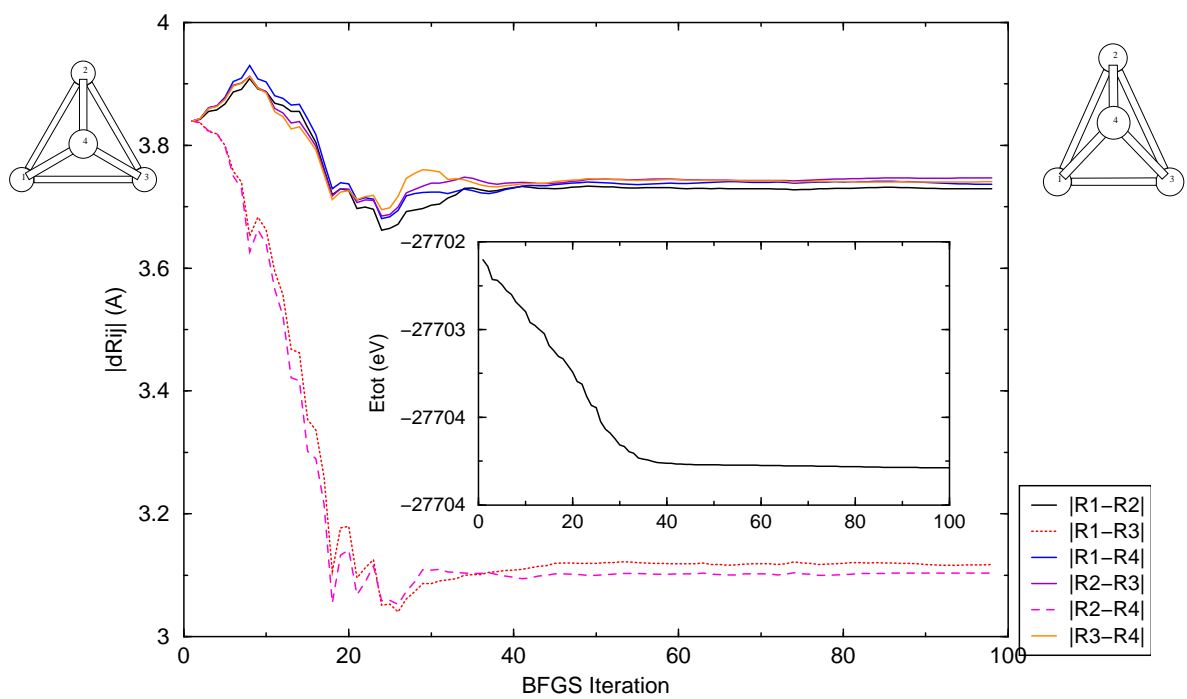}} \par}

\caption{\label{fig: si255vc_BFGS}Relaxation of vacancy using a standard
BFGS minimizer. The 6 distances between the 4 silicon atoms surrounding
the vacancy are shown. This clearly shows the change in symmetry around
the defect, with the initial and final states of the first shell of
atoms around the vacancy shown. The atoms are numbered as in the sketches.
In the initial state, all bond lengths are equal and the defect has
\protect\( T_{d}\protect \)-point symmetry, whereas in the final
relaxed state of the first shell of atoms, there are 4 equal, longer
bond lengths and two equal, shorter bond lengths which therefore corresponds
to \protect\( D_{2d}\protect \) -point symmetry.}

Also shown in the inset is the convergence of the total energy of
the system as the relaxation proceeds. The relaxation lowers the energy
of the system by \( 1.186\: eV \).
\end{figure}

\begin{figure}
{\centering \resizebox*{0.9\columnwidth}{!}{\rotatebox{270}{\includegraphics{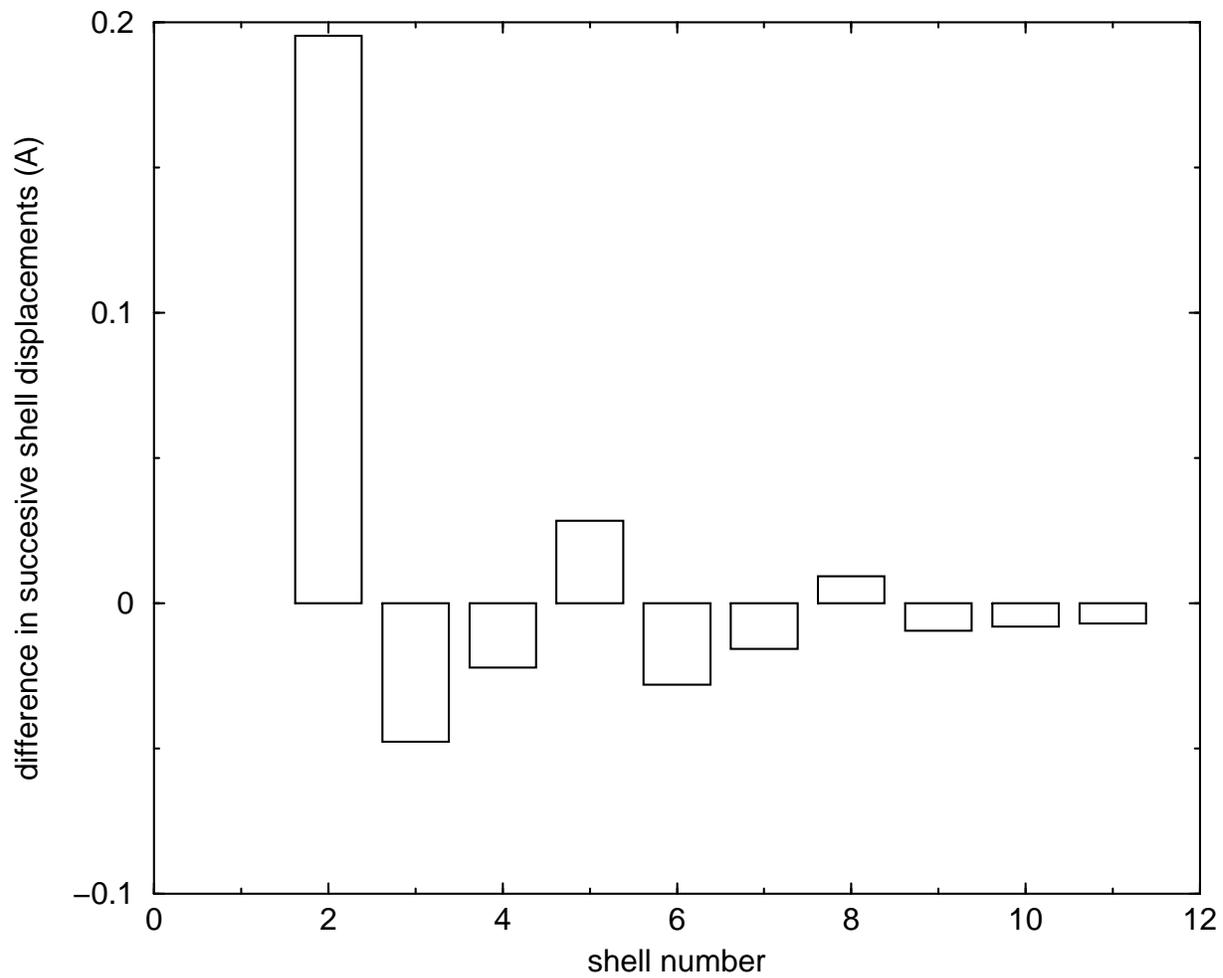}}} \par}

\caption{\label{fig: si255vc_shells}Convergence of the ionic relaxation of
successive shells of atoms across the supercell.}
\end{figure}

For the neutral silicon vacancy we have therefore established the
necessary parameters to achieve an accurate energy surface, and so
only now do we relax the vacancy in the 255 atom BCC supercell, without
using any symmetrization of the electronic parameters (wavefunction,
charge density, forces, etc) at any stage. This is necessary to ensure
that any symmetry in the relaxed structure is spontaneous and not
imposed from the initial conditions. Tight convergence tolerances
are imposed - namely that at convergence the RMS force be less than
\( 0.001\: eV/ \)\AA, the RMS displacement be less than \( 0.0001 \)
\AA{ } per iteration, and that the energy difference per iteration
be less than \( 0.00001\: eV/atom \). The results of such a calculation
are shown in figure \ref{fig: si255vc_BFGS}. This calculation is
also repeated for different random perturbations of the atoms in the
first shell surrounding the vacancy, to ensure that the same minimum
structure is reached each time. To test that the atomic relaxation
is contained within the supercell, we calculate the relative displacements
of the successive shells of atoms surrounding the vacancy as shown
in figure \ref{fig: si255vc_shells}. From this we can see that shells
9-11 are essentially unchanged (where shell 12 is the half-way point
in the supercell), and so we conclude that the ionic relaxation is
fully contained within the finite size of the supercell.

\section{Results}

\label{sec:results}

We now summarize our results for the neutral silicon vacancy.

It was found that when converging the electronic structure of the
unrelaxed vacancy, that BCC supercells gave superior finite-size supercell
convergence, and that a 256/255 atom BCC supercell was required to
get satisfactory convergence which was confirmed against an 864/863
atom calculation. Remarkably, the 32/31 atom BCC supercell gave an
unrelaxed vacancy formation energy which was close to the infinite
supercell size limit than that of the 250/249 FCC supercell calculation.
This is attributable to the interaction of supercell symmetry and
the symmetry of the underlying silicon lattice.

As a measure of the reliability of the \emph{ab initio} scheme used,
we repeated the calculation of the unrelaxed vacancy formation energy
in the 32/31 atom BCC supercell using different exchange-correlation
functionals and different pseudopotentials. The results are summarised
in table \ref{table: DFT errors}. It has been found many times before
that there is a general tendency for LDA-DFT calculations to over-bind,
and GGA-DFT calculations to under-bind. Hence, we conclude that a
worst-case error estimate for our \emph{ab initio} scheme is \( \pm 0.02\: eV \),
but a more likely error estimate is \( \pm 0.01\: eV \). It can also
be seen that the systematic convergence studies as presented above,
such as the Brillouin zone sampling, can make a more significant change
than changing the exchange-correlation functional at a given set of
parameters (e.g. going from \( q=2 \) to \( q=4 \) reduces the formation
energy by \( \sim 0.08\: eV \)). Of course, as noted in section \ref{sec:intro},
a more thorough comparison would be between DFT and QMC calculations,
but at present there are no available QMC data to compare against.

The 255 atom BCC supercell was then used to relax the defect structure,
and it was demonstrated that this relaxation was fully contained within
the supercell. This relaxation reduced the total energy of the system
by \( 1.186\: eV \) and from the observed bond lengths of the 4 atoms
in the first shell surrounding the vacancy, we can see that the final
relaxed structure has spontaneously achieved the \( D_{2d} \) -point
symmetry, with a final volume (as given by equation \ref{eqn: volume of vacancy})
that is reduced from the unrelaxed vacancy by \( -27\: \% \). The
relaxed defect formation energy is therefore estimated as \( 3.17\pm 0.01\: eV \)
(where the error estimate is that due to the the \emph{ab initio}
scheme used - the convergence error estimate is an order of magnitude
smaller). The final parameters used in the calculation and the final
result for the structure of the defect are summarized in table \ref{table: final results}.

\section{Conclusions}

\label{sec:conc} We have presented a systematic methodology for the
accurate calculation of defect structures in supercells. Various potential
pitfalls have been highlighted, and it has been demonstrated how to
systematically reduce each source of error in the various convergence
parameters, to better than the inherent accuracy of the \emph{ab initio}
method used.

As an example of the methodology, the single neutral vacancy in silicon
has been treated. This has been extensively studied in the past, but
with many different answers presented in the literature. The various
sources of scatter in previous results have been discussed, such as
problems with too small a basis-set size leading to a spurious outwards
relaxation as seen in the earlier studies, and problems caused by
under-convergence of the Brillouin zone sampling leading to inaccurate
forces in some more recent studies. The use of offset grids has been
shown to be very useful in accelerating the convergence of the Brillouin
zone sampling. A new effect, that of supercell symmetry, has been
identified, and a consistent treatment of this has been shown to be
crucial to understanding the systematic effects of increasing the
supercell size. This has resulted in great difficulty in the past
with identifying the convergence trends with increasing supercell
size, and it is shown herein that the best systematic way to treat
this effect is to consider the defect density for each different supercell
symmetry separately. Therefore it is believed that this work presents
the best converged calculation of the silicon vacancy to date.

\begin{acknowledgments}
Financial support for MIJP was provided by the Lloyds of London Tercentenary
Foundation. The calculations were performed on the Silicon Graphics
Origin 2000 at CSAR, and on the Hitachi SR2201 at the University of
Cambridge HPCF.
\end{acknowledgments}

\bibliographystyle{apsrev}

\begin{thebibliography}{43}
\expandafter\ifx\csname natexlab\endcsname\relax\def\natexlab#1{#1}\fi
\expandafter\ifx\csname bibnamefont\endcsname\relax
  \def\bibnamefont#1{#1}\fi
\expandafter\ifx\csname bibfnamefont\endcsname\relax
  \def\bibfnamefont#1{#1}\fi
\expandafter\ifx\csname citenamefont\endcsname\relax
  \def\citenamefont#1{#1}\fi
\expandafter\ifx\csname url\endcsname\relax
  \def\url#1{\texttt{#1}}\fi
\expandafter\ifx\csname urlprefix\endcsname\relax\def\urlprefix{URL }\fi
\providecommand{\bibinfo}[2]{#2}
\providecommand{\eprint}[2][]{\url{#2}}

\bibitem[{\citenamefont{Jones and Gunnarsson}(1989)}]{JonesG89}
\bibinfo{author}{\bibfnamefont{R.~O.} \bibnamefont{Jones}} \bibnamefont{and}
  \bibinfo{author}{\bibfnamefont{O.}~\bibnamefont{Gunnarsson}},
  \bibinfo{journal}{Rev. Mod. Phys.} \textbf{\bibinfo{volume}{61}},
  \bibinfo{pages}{689} (\bibinfo{year}{1989}).

\bibitem[{\citenamefont{Payne et~al.}(1992)\citenamefont{Payne, Teter, Allan,
  Arias, and Joannopoulos}}]{PayneTAAJ92}
\bibinfo{author}{\bibfnamefont{M.~C.} \bibnamefont{Payne}},
  \bibinfo{author}{\bibfnamefont{M.~P.} \bibnamefont{Teter}},
  \bibinfo{author}{\bibfnamefont{D.~C.} \bibnamefont{Allan}},
  \bibinfo{author}{\bibfnamefont{T.~A.} \bibnamefont{Arias}}, \bibnamefont{and}
  \bibinfo{author}{\bibfnamefont{J.~D.} \bibnamefont{Joannopoulos}},
  \bibinfo{journal}{Rev. Mod. Phys.} \textbf{\bibinfo{volume}{64}},
  \bibinfo{pages}{1045} (\bibinfo{year}{1992}).

\bibitem[{\citenamefont{Foulkes et~al.}(2001)\citenamefont{Foulkes, Mitas,
  Needs, and Rajagopal}}]{FoulkesMNR01}
\bibinfo{author}{\bibfnamefont{W.~M.~C.} \bibnamefont{Foulkes}},
  \bibinfo{author}{\bibfnamefont{L.}~\bibnamefont{Mitas}},
  \bibinfo{author}{\bibfnamefont{R.~J.} \bibnamefont{Needs}}, \bibnamefont{and}
  \bibinfo{author}{\bibfnamefont{G.}~\bibnamefont{Rajagopal}},
  \bibinfo{journal}{Rev. Mod. Phys.} \textbf{\bibinfo{volume}{73}},
  \bibinfo{pages}{33} (\bibinfo{year}{2001}).

\bibitem[{\citenamefont{Watkins}(1986)}]{Watkins86}
\bibinfo{author}{\bibfnamefont{G.}~\bibnamefont{Watkins}}, in
  \emph{\bibinfo{booktitle}{Deep Centres in Semiconductors}}, edited by
  \bibinfo{editor}{\bibfnamefont{S.}~\bibnamefont{Pantiledes}}
  (\bibinfo{publisher}{Gordon and Breach}, \bibinfo{address}{New York},
  \bibinfo{year}{1986}), p. \bibinfo{pages}{147}.

\bibitem[{\citenamefont{Watkins}(1976)}]{Watkins76}
\bibinfo{author}{\bibfnamefont{G.}~\bibnamefont{Watkins}}, in
  \emph{\bibinfo{booktitle}{Defects and Their Structure in Non-metallic
  Solids}}, edited by \bibinfo{editor}{\bibfnamefont{H.}~\bibnamefont{B.}}
  \bibnamefont{and} \bibinfo{editor}{\bibfnamefont{H.}~\bibnamefont{A.E.}}
  (\bibinfo{publisher}{Plenum}, \bibinfo{address}{New York},
  \bibinfo{year}{1976}), p. \bibinfo{pages}{203}.

\bibitem[{\citenamefont{Sprenger et~al.}(1983)\citenamefont{Sprenger, Muller,
  and Ammerlaan}}]{SprengerMA83}
\bibinfo{author}{\bibfnamefont{M.}~\bibnamefont{Sprenger}},
  \bibinfo{author}{\bibfnamefont{S.~H.} \bibnamefont{Muller}},
  \bibnamefont{and} \bibinfo{author}{\bibfnamefont{C.~A.~J.}
  \bibnamefont{Ammerlaan}}, \bibinfo{journal}{Physica B}
  \textbf{\bibinfo{volume}{116}}, \bibinfo{pages}{224} (\bibinfo{year}{1983}).

\bibitem[{\citenamefont{Sprenger et~al.}(1987)\citenamefont{Sprenger, Muller,
  Sieverts, and Ammerlaan}}]{SprengerMSA87}
\bibinfo{author}{\bibfnamefont{M.}~\bibnamefont{Sprenger}},
  \bibinfo{author}{\bibfnamefont{S.~H.} \bibnamefont{Muller}},
  \bibinfo{author}{\bibfnamefont{E.~G.} \bibnamefont{Sieverts}},
  \bibnamefont{and} \bibinfo{author}{\bibfnamefont{C.~A.~J.}
  \bibnamefont{Ammerlaan}}, \bibinfo{journal}{Phys. Rev. B}
  \textbf{\bibinfo{volume}{35}}, \bibinfo{pages}{1566} (\bibinfo{year}{1987}).

\bibitem[{\citenamefont{Samara}(1987)}]{Samara87}
\bibinfo{author}{\bibfnamefont{G.~A.} \bibnamefont{Samara}},
  \bibinfo{journal}{Phys. Rev. B} \textbf{\bibinfo{volume}{36}},
  \bibinfo{pages}{4841} (\bibinfo{year}{1987}).

\bibitem[{\citenamefont{Samara}(1988)}]{Samara88}
\bibinfo{author}{\bibfnamefont{G.~A.} \bibnamefont{Samara}},
  \bibinfo{journal}{Phys. Rev. B} \textbf{\bibinfo{volume}{37}},
  \bibinfo{pages}{8523} (\bibinfo{year}{1988}).

\bibitem[{\citenamefont{Samara}(1989{\natexlab{a}})}]{Samara89a}
\bibinfo{author}{\bibfnamefont{G.~A.} \bibnamefont{Samara}},
  \bibinfo{journal}{Phys. Rev. B} \textbf{\bibinfo{volume}{39}},
  \bibinfo{pages}{11001} (\bibinfo{year}{1989}{\natexlab{a}}).

\bibitem[{\citenamefont{Samara}(1989{\natexlab{b}})}]{Samara89b}
\bibinfo{author}{\bibfnamefont{G.~A.} \bibnamefont{Samara}},
  \bibinfo{journal}{Phys. Rev. B} \textbf{\bibinfo{volume}{39}},
  \bibinfo{pages}{12764} (\bibinfo{year}{1989}{\natexlab{b}}).

\bibitem[{\citenamefont{M{\'a}kinen et~al.}(1989)\citenamefont{M{\'a}kinen,
  Corbel, Hautoj{\'a}rvi, Moser, and Pierre}}]{MakinenCHMP89}
\bibinfo{author}{\bibfnamefont{J.}~\bibnamefont{M{\'a}kinen}},
  \bibinfo{author}{\bibfnamefont{C.}~\bibnamefont{Corbel}},
  \bibinfo{author}{\bibfnamefont{P.}~\bibnamefont{Hautoj{\'a}rvi}},
  \bibinfo{author}{\bibfnamefont{P.}~\bibnamefont{Moser}}, \bibnamefont{and}
  \bibinfo{author}{\bibfnamefont{F.}~\bibnamefont{Pierre}},
  \bibinfo{journal}{Phys. Rev. B} \textbf{\bibinfo{volume}{39}},
  \bibinfo{pages}{10162} (\bibinfo{year}{1989}).

\bibitem[{\citenamefont{M{\'a}kinen et~al.}(1992)\citenamefont{M{\'a}kinen,
  Hautoj{\'a}rvi, and Corbel}}]{MakinenHC92}
\bibinfo{author}{\bibfnamefont{J.}~\bibnamefont{M{\'a}kinen}},
  \bibinfo{author}{\bibfnamefont{P.}~\bibnamefont{Hautoj{\'a}rvi}},
  \bibnamefont{and} \bibinfo{author}{\bibfnamefont{C.}~\bibnamefont{Corbel}},
  \bibinfo{journal}{J. Phys.-Condes. Matter} \textbf{\bibinfo{volume}{4}},
  \bibinfo{pages}{5137} (\bibinfo{year}{1992}).

\bibitem[{\citenamefont{Scheffler}(1982)}]{Scheffler82}
\bibinfo{author}{\bibfnamefont{M.}~\bibnamefont{Scheffler}},
  \bibinfo{journal}{Fertkorperprobleme-Adv. Solid State Phys.}
  \textbf{\bibinfo{volume}{22}}, \bibinfo{pages}{115} (\bibinfo{year}{1982}).

\bibitem[{\citenamefont{Gunnarsson et~al.}(1983)\citenamefont{Gunnarsson,
  Jepsen, and Andersen}}]{GunnarssonJA83}
\bibinfo{author}{\bibfnamefont{O.}~\bibnamefont{Gunnarsson}},
  \bibinfo{author}{\bibfnamefont{O.}~\bibnamefont{Jepsen}}, \bibnamefont{and}
  \bibinfo{author}{\bibfnamefont{O.~K.} \bibnamefont{Andersen}},
  \bibinfo{journal}{Phys. Rev. B} \textbf{\bibinfo{volume}{27}},
  \bibinfo{pages}{7144} (\bibinfo{year}{1983}).

\bibitem[{\citenamefont{Scheffler et~al.}(1985)\citenamefont{Scheffler,
  Vigneron, and Bachelet}}]{SchefflerV85}
\bibinfo{author}{\bibfnamefont{M.}~\bibnamefont{Scheffler}},
  \bibinfo{author}{\bibfnamefont{J.~P.} \bibnamefont{Vigneron}},
  \bibnamefont{and} \bibinfo{author}{\bibfnamefont{G.}~\bibnamefont{Bachelet}},
  \bibinfo{journal}{Phys. Rev. B} \textbf{\bibinfo{volume}{31}},
  \bibinfo{pages}{6541} (\bibinfo{year}{1985}).

\bibitem[{\citenamefont{Wang et~al.}(1991)\citenamefont{Wang, Chan, and
  Ho}}]{WangCH91}
\bibinfo{author}{\bibfnamefont{C.~Z.} \bibnamefont{Wang}},
  \bibinfo{author}{\bibfnamefont{C.~T.} \bibnamefont{Chan}}, \bibnamefont{and}
  \bibinfo{author}{\bibfnamefont{K.~M.} \bibnamefont{Ho}},
  \bibinfo{journal}{Phys. Rev. Lett.} \textbf{\bibinfo{volume}{66}},
  \bibinfo{pages}{189} (\bibinfo{year}{1991}).

\bibitem[{\citenamefont{Munro and Wales}(1999)}]{MunroW99}
\bibinfo{author}{\bibfnamefont{L.~J.} \bibnamefont{Munro}} \bibnamefont{and}
  \bibinfo{author}{\bibfnamefont{D.~J.} \bibnamefont{Wales}},
  \bibinfo{journal}{Phys. Rev. B} \textbf{\bibinfo{volume}{59}},
  \bibinfo{pages}{3969} (\bibinfo{year}{1999}).

\bibitem[{\citenamefont{Antonelli and Bernholc}(1989)}]{AntonelliB89}
\bibinfo{author}{\bibfnamefont{A.}~\bibnamefont{Antonelli}} \bibnamefont{and}
  \bibinfo{author}{\bibfnamefont{J.}~\bibnamefont{Bernholc}},
  \bibinfo{journal}{Phys. Rev. B} \textbf{\bibinfo{volume}{40}},
  \bibinfo{pages}{10643} (\bibinfo{year}{1989}).

\bibitem[{\citenamefont{Sugino and Oshiyama}(1992)}]{SuginoO92}
\bibinfo{author}{\bibfnamefont{O.}~\bibnamefont{Sugino}} \bibnamefont{and}
  \bibinfo{author}{\bibfnamefont{A.}~\bibnamefont{Oshiyama}},
  \bibinfo{journal}{Phys. Rev. Lett.} \textbf{\bibinfo{volume}{68}},
  \bibinfo{pages}{1858} (\bibinfo{year}{1992}).

\bibitem[{\citenamefont{Bl{\"o}chl et~al.}(1993)\citenamefont{Bl{\"o}chl,
  Smargiassi, Car, Laks, Andreoni, and Pantelides}}]{BlochlSCLAP93}
\bibinfo{author}{\bibfnamefont{P.~E.} \bibnamefont{Bl{\"o}chl}},
  \bibinfo{author}{\bibfnamefont{E.}~\bibnamefont{Smargiassi}},
  \bibinfo{author}{\bibfnamefont{R.}~\bibnamefont{Car}},
  \bibinfo{author}{\bibfnamefont{D.~B.} \bibnamefont{Laks}},
  \bibinfo{author}{\bibfnamefont{W.}~\bibnamefont{Andreoni}}, \bibnamefont{and}
  \bibinfo{author}{\bibfnamefont{S.~T.} \bibnamefont{Pantelides}},
  \bibinfo{journal}{Phys. Rev. Lett.} \textbf{\bibinfo{volume}{70}},
  \bibinfo{pages}{2435} (\bibinfo{year}{1993}).

\bibitem[{\citenamefont{Ramamoorthy and Pantelides}(1996)}]{RamamoorthyP96}
\bibinfo{author}{\bibfnamefont{M.}~\bibnamefont{Ramamoorthy}} \bibnamefont{and}
  \bibinfo{author}{\bibfnamefont{S.~T.} \bibnamefont{Pantelides}},
  \bibinfo{journal}{Phys. Rev. Lett.} \textbf{\bibinfo{volume}{76}},
  \bibinfo{pages}{4753} (\bibinfo{year}{1996}).

\bibitem[{\citenamefont{Pankratov et~al.}(1997)\citenamefont{Pankratov, Huang,
  Diazdelarubia, and Mailhiot}}]{PankratovHDM97}
\bibinfo{author}{\bibfnamefont{O.}~\bibnamefont{Pankratov}},
  \bibinfo{author}{\bibfnamefont{H.~C.} \bibnamefont{Huang}},
  \bibinfo{author}{\bibfnamefont{T.}~\bibnamefont{Diazdelarubia}},
  \bibnamefont{and} \bibinfo{author}{\bibfnamefont{C.}~\bibnamefont{Mailhiot}},
  \bibinfo{journal}{Phys. Rev. B} \textbf{\bibinfo{volume}{56}},
  \bibinfo{pages}{13172} (\bibinfo{year}{1997}).

\bibitem[{\citenamefont{Zywietz et~al.}(1997)\citenamefont{Zywietz,
  Furthm{\"u}ller, and Bechstedt}}]{ZywietzFB97}
\bibinfo{author}{\bibfnamefont{A.}~\bibnamefont{Zywietz}},
  \bibinfo{author}{\bibfnamefont{J.}~\bibnamefont{Furthm{\"u}ller}},
  \bibnamefont{and}
  \bibinfo{author}{\bibfnamefont{F.}~\bibnamefont{Bechstedt}},
  \bibinfo{journal}{Mater. Sci. Forum} \textbf{\bibinfo{volume}{258-2}},
  \bibinfo{pages}{653} (\bibinfo{year}{1997}).

\bibitem[{\citenamefont{Puska et~al.}(1998)\citenamefont{Puska, Poykko, Pesola,
  and Nieminen}}]{PuskaPPN98}
\bibinfo{author}{\bibfnamefont{M.~J.} \bibnamefont{Puska}},
  \bibinfo{author}{\bibfnamefont{S.}~\bibnamefont{Poykko}},
  \bibinfo{author}{\bibfnamefont{M.}~\bibnamefont{Pesola}}, \bibnamefont{and}
  \bibinfo{author}{\bibfnamefont{R.~M.} \bibnamefont{Nieminen}},
  \bibinfo{journal}{Phys. Rev. B} \textbf{\bibinfo{volume}{58}},
  \bibinfo{pages}{1318} (\bibinfo{year}{1998}).

\bibitem[{\citenamefont{Ogut et~al.}(1997)\citenamefont{Ogut, Kim, and
  Chelikowsky}}]{OgutKC97}
\bibinfo{author}{\bibfnamefont{S.}~\bibnamefont{Ogut}},
  \bibinfo{author}{\bibfnamefont{H.}~\bibnamefont{Kim}}, \bibnamefont{and}
  \bibinfo{author}{\bibfnamefont{J.~R.} \bibnamefont{Chelikowsky}},
  \bibinfo{journal}{Phys. Rev. B} \textbf{\bibinfo{volume}{56}},
  \bibinfo{pages}{11353} (\bibinfo{year}{1997}).

\bibitem[{\citenamefont{{Molecular Simulations Inc.}}(1999)}]{MSI99}
\bibinfo{author}{\bibnamefont{{Molecular Simulations Inc.}}},
  \emph{\bibinfo{title}{{CASTEP} 4.2 {Academic} version, licensed under the
  {UKCP-MSI} Agreement}} (\bibinfo{year}{1999}).

\bibitem[{\citenamefont{Perdew}(1991)}]{Perdew91}
\bibinfo{author}{\bibfnamefont{J.~P.} \bibnamefont{Perdew}}, in
  \emph{\bibinfo{booktitle}{Electronic Structure of Solids '91}}, edited by
  \bibinfo{editor}{\bibfnamefont{P.}~\bibnamefont{Zeische}} \bibnamefont{and}
  \bibinfo{editor}{\bibfnamefont{H.}~\bibnamefont{Eschrig}}
  (\bibinfo{publisher}{Akademie Verlag}, \bibinfo{address}{Berlin},
  \bibinfo{year}{1991}).

\bibitem[{\citenamefont{Vanderbilt}(1990)}]{Vanderbilt90}
\bibinfo{author}{\bibfnamefont{D.}~\bibnamefont{Vanderbilt}},
  \bibinfo{journal}{Phys. Rev. B} \textbf{\bibinfo{volume}{41}},
  \bibinfo{pages}{7892} (\bibinfo{year}{1990}).

\bibitem[{\citenamefont{Ceperley and Alder}(1980)}]{CeperleyA80}
\bibinfo{author}{\bibfnamefont{D.~M.} \bibnamefont{Ceperley}} \bibnamefont{and}
  \bibinfo{author}{\bibfnamefont{B.~J.} \bibnamefont{Alder}},
  \bibinfo{journal}{Phys. Rev. Lett.} \textbf{\bibinfo{volume}{45}},
  \bibinfo{pages}{566} (\bibinfo{year}{1980}).

\bibitem[{\citenamefont{Perdew et~al.}(1996)\citenamefont{Perdew, Burke, and
  Ernzerhof}}]{PerdewBE96}
\bibinfo{author}{\bibfnamefont{J.~P.} \bibnamefont{Perdew}},
  \bibinfo{author}{\bibfnamefont{K.}~\bibnamefont{Burke}}, \bibnamefont{and}
  \bibinfo{author}{\bibfnamefont{M.}~\bibnamefont{Ernzerhof}},
  \bibinfo{journal}{Phys. Rev. Lett.} \textbf{\bibinfo{volume}{77}},
  \bibinfo{pages}{3865} (\bibinfo{year}{1996}).

\bibitem[{\citenamefont{Bachelet et~al.}(1982)\citenamefont{Bachelet, Hamann,
  and Schluter}}]{BacheletHS82}
\bibinfo{author}{\bibfnamefont{G.~B.} \bibnamefont{Bachelet}},
  \bibinfo{author}{\bibfnamefont{D.~R.} \bibnamefont{Hamann}},
  \bibnamefont{and} \bibinfo{author}{\bibfnamefont{M.}~\bibnamefont{Schluter}},
  \bibinfo{journal}{Phys. Rev. B} \textbf{\bibinfo{volume}{26}},
  \bibinfo{pages}{4199} (\bibinfo{year}{1982}).

\bibitem[{\citenamefont{Hamann}(1989)}]{Hamann89}
\bibinfo{author}{\bibfnamefont{D.~R.} \bibnamefont{Hamann}},
  \bibinfo{journal}{Phys. Rev. B} \textbf{\bibinfo{volume}{40}},
  \bibinfo{pages}{2980} (\bibinfo{year}{1989}).

\bibitem[{\citenamefont{Kleinman and Bylander}(1982)}]{KleinmanB82}
\bibinfo{author}{\bibfnamefont{L.}~\bibnamefont{Kleinman}} \bibnamefont{and}
  \bibinfo{author}{\bibfnamefont{D.~M.} \bibnamefont{Bylander}},
  \bibinfo{journal}{Phys. Rev. Lett.} \textbf{\bibinfo{volume}{48}},
  \bibinfo{pages}{1425} (\bibinfo{year}{1982}).

\bibitem[{\citenamefont{Monkhorst and Pack}(1976)}]{MonkhorstP76}
\bibinfo{author}{\bibfnamefont{H.~J.} \bibnamefont{Monkhorst}}
  \bibnamefont{and} \bibinfo{author}{\bibfnamefont{J.~D.} \bibnamefont{Pack}},
  \bibinfo{journal}{Phys. Rev. B} \textbf{\bibinfo{volume}{13}},
  \bibinfo{pages}{5188} (\bibinfo{year}{1976}).

\bibitem[{\citenamefont{Moreno and Soler}(1992)}]{MorenoS92}
\bibinfo{author}{\bibfnamefont{J.}~\bibnamefont{Moreno}} \bibnamefont{and}
  \bibinfo{author}{\bibfnamefont{J.~M.} \bibnamefont{Soler}},
  \bibinfo{journal}{Phys. Rev. B} \textbf{\bibinfo{volume}{45}},
  \bibinfo{pages}{13891} (\bibinfo{year}{1992}).

\bibitem[{\citenamefont{Makov et~al.}(1996)\citenamefont{Makov, Shah, and
  Payne}}]{MakovSP96}
\bibinfo{author}{\bibfnamefont{G.}~\bibnamefont{Makov}},
  \bibinfo{author}{\bibfnamefont{R.}~\bibnamefont{Shah}}, \bibnamefont{and}
  \bibinfo{author}{\bibfnamefont{M.~C.} \bibnamefont{Payne}},
  \bibinfo{journal}{Phys. Rev. B} \textbf{\bibinfo{volume}{53}},
  \bibinfo{pages}{15513} (\bibinfo{year}{1996}).

\bibitem[{\citenamefont{Ashcroft and Mermin}(1976)}]{AshcroftM76_10}
\bibinfo{author}{\bibfnamefont{N.~W.} \bibnamefont{Ashcroft}} \bibnamefont{and}
  \bibinfo{author}{\bibfnamefont{N.~D.} \bibnamefont{Mermin}},
  \emph{\bibinfo{title}{Solid State Physics}} (\bibinfo{publisher}{Saunders
  College}, \bibinfo{address}{Philadelphia}, \bibinfo{year}{1976}),
  chap.~\bibinfo{chapter}{10}.

\bibitem[{\citenamefont{Springborg}(2000)}]{Springborg00}
\bibinfo{author}{\bibfnamefont{M.}~\bibnamefont{Springborg}},
  \emph{\bibinfo{title}{Methods of Electronic-Structure Calculations}}
  (\bibinfo{publisher}{John Wiley and Sons Ltd}, \bibinfo{address}{Chichester,
  England}, \bibinfo{year}{2000}).

\bibitem[{\citenamefont{Makov and Payne}(1995)}]{MakovP95}
\bibinfo{author}{\bibfnamefont{G.}~\bibnamefont{Makov}} \bibnamefont{and}
  \bibinfo{author}{\bibfnamefont{M.~C.} \bibnamefont{Payne}},
  \bibinfo{journal}{Phys. Rev. B} \textbf{\bibinfo{volume}{51}},
  \bibinfo{pages}{4014} (\bibinfo{year}{1995}).

\bibitem[{\citenamefont{Kirkpatrick et~al.}(1983)\citenamefont{Kirkpatrick,
  Gelatt, and Vecchi}}]{KirkpatrickGV83}
\bibinfo{author}{\bibfnamefont{S.}~\bibnamefont{Kirkpatrick}},
  \bibinfo{author}{\bibfnamefont{C.~D.} \bibnamefont{Gelatt}},
  \bibnamefont{and} \bibinfo{author}{\bibfnamefont{M.~P.}
  \bibnamefont{Vecchi}}, \bibinfo{journal}{Science}
  \textbf{\bibinfo{volume}{220}}, \bibinfo{pages}{671} (\bibinfo{year}{1983}).

\bibitem[{\citenamefont{Car and Parrinello}(1985)}]{CarP85}
\bibinfo{author}{\bibfnamefont{R.}~\bibnamefont{Car}} \bibnamefont{and}
  \bibinfo{author}{\bibfnamefont{M.}~\bibnamefont{Parrinello}},
  \bibinfo{journal}{Phys. Rev. Lett.} \textbf{\bibinfo{volume}{55}},
  \bibinfo{pages}{2471} (\bibinfo{year}{1985}).

\bibitem[{\citenamefont{Frenkel and Smit}(1996)}]{FrenkelS96_7}
\bibinfo{author}{\bibfnamefont{D.}~\bibnamefont{Frenkel}} \bibnamefont{and}
  \bibinfo{author}{\bibfnamefont{B.}~\bibnamefont{Smit}},
  \emph{\bibinfo{title}{Understanding Molecular Simulations}}
  (\bibinfo{publisher}{Academic Press}, \bibinfo{address}{San Diego},
  \bibinfo{year}{1996}), chap.~\bibinfo{chapter}{7}.

\end{thebibliography}

\clearpage

\begin{table*}

\caption{\label{table: simple supercells, no offsets}Effect of increasing
the Monkhorst-Pack grid parameter \protect\( q\protect \) on the
number of symmetrized points in the grid (\protect\( N_{s}\protect \))
and the (squared) radius of exact integration (\protect\( R^{2}\protect \))
in units of reciprocal lattice vectors, for three different supercell
symmetries.}

\begin{tabular}{|c|c|c|c|c|c|c|}
\hline 
Monkhorst-Pack&
\multicolumn{2}{|c|}{Simple Cubic }&
\multicolumn{2}{|c|}{Body-Centred Cubic }&
\multicolumn{2}{|c|}{Face-Centred Cubic }\\
\hline 
 \( q \)&
 \( N_{s} \)&
 \( R^{2} \)&
 \( N_{s} \)&
 \( R^{2} \)&
 \( N_{s} \)&
 \( R^{2} \)\\
\hline
1&
 1&
 1.0&
 1&
 0.75&
 1&
 0.5\\
\hline
2&
 1&
 4.0&
 2&
 3.0&
 2&
 4.0\\
\hline
3&
 4&
 9.0&
 4&
 6.75&
 4&
 4.5\\
\hline
4&
 4&
 16.0&
 6&
 12.0&
 10&
 16.0 \\
\hline
\end{tabular}
\end{table*}

\begin{table*}

\caption{\label{table: simple supercells, with offsets}Effect of optimal
offset \protect\( \mathbf{k}_{0}\protect \) on maximizing the efficiency
of the Brillouin zone integration for three different supercell symmetries
with increasing values of the Monkhorst-Pack grid parameter \protect\( q\protect \).
The number of symmetrized points in the grid (\protect\( N_{s}\protect \))
and the (squared) radius of exact integration (\protect\( R^{2}\protect \))
in units of reciprocal lattice vectors is given for each optimal offset.}

\begin{tabular}{|c|c|c|c|c|c|c|c|c|c|}
\hline 
Monkhorst-Pack&
\multicolumn{3}{c|}{Simple Cubic}&
\multicolumn{3}{|c|}{Body-Centred Cubic }&
\multicolumn{3}{|c|}{Face-Centred Cubic }\\
\hline 
 \( q \)&
 \( \mathbf{k}_{0} \)&
 \( N_{s} \)&
 \( R^{2} \)&
 \( \mathbf{k}_{0} \)&
 \( N_{s} \)&
 \( R^{2} \)&
 \( \mathbf{k}_{0} \)&
 \( N_{s} \)&
 \( R^{2} \)\\
\hline
1&
 \( \left( \frac{1}{4},\frac{1}{4},\frac{1}{4}\right)  \)&
 1&
 4.0&
 \( \left( 0,\frac{1}{4},\frac{1}{2}\right)  \)&
 1&
 2.0&
 \( \left( 0,\frac{1}{2},\frac{1}{2}\right)  \)&
 1&
 1.0\\
\hline
2&
 \( \left( \frac{1}{8},\frac{1}{8},\frac{1}{8}\right)  \)&
 3&
 16.0&
 \( \left( \frac{1}{4},\frac{1}{4},\frac{1}{4}\right)  \)&
 2&
 4.0&
 \( \left( 0,0,0\right)  \)&
 2&
 4.0\\
\hline
3&
 \( \left( \frac{1}{4},0,\frac{1}{2}\right)  \)&
 8&
 18.0&
 \( \left( \frac{1}{2},\frac{1}{2},\frac{1}{2}\right)  \)&
 5&
 9.0&
 \( \left( \frac{1}{2},\frac{1}{2},\frac{1}{2}\right)  \)&
 6&
 9.0\\
\hline
4&
 \( \left( \frac{1}{16},\frac{1}{16},\frac{1}{16}\right)  \)&
 20&
 64.0&
 \( \left( \frac{1}{8},\frac{1}{8},\frac{1}{8}\right)  \)&
 8&
 16.0 &
 \( \left( 0,0,0\right)  \)&
 10&
 16.0 \\
\hline
\end{tabular}
\end{table*}

\begin{table}

\caption{\label{table: supercell list}List of all supercells considered with
corresponding supercell symmetry. Also listed is the converged value
of \protect\( q\protect \) (Monkhorst-Pack grid parameter) used in
the calculation, the Monkhorst-Pack grid offset \textbf{\protect\( \mathbf{k}_{0}\protect \)}
used, and the corresponding Brillouin zone sampling density.}

\begin{tabular}{|c|c|c|c|c|}
\hline 
\( N \)&
 Symmetry&
 \( q \)&
\textbf{\( \mathbf{k}_{0} \)}&
BZ density (\AA\( ^{-1} \))\\
\hline
2&
 FCC&
 8&
\( \left( 0,0,0\right)  \)&
 0.040\\
\hline
8&
 SC&
 6&
\( \left( 0,0,0\right)  \)&
 0.031\\
\hline
16&
 FCC&
 4&
\( \left( 0,0,0\right)  \)&
 0.040\\
\hline
32&
 BCC&
 4&
\( \left( \frac{1}{8},\frac{1}{8},\frac{1}{8}\right)  \)&
 0.033\\
\hline
54&
 FCC&
 3&
\( \left( \frac{1}{2},\frac{1}{2},\frac{1}{2}\right)  \)&
 0.036\\
\hline
64&
 SC&
 3&
\( \left( 0,0,0\right)  \)&
 0.031\\
\hline
128&
 FCC&
 2&
\( \left( 0,0,0\right)  \)&
 0.040\\
\hline
216&
 SC&
 2&
\( \left( 0,0,0\right)  \)&
 0.031\\
\hline
250&
 FCC&
 2&
\( \left( 0,0,0\right)  \)&
 0.032\\
\hline
256&
 BCC&
 2&
\( \left( \frac{1}{4},\frac{1}{4},\frac{1}{4}\right)  \)&
 0.033\\
\hline
\end{tabular}
\end{table}

\begin{table}

\caption{\label{table: DFT errors}The unrelaxed vacancy formation energy
for the 32/31 atom BCC supercell, with different exchange-correlation
functionals and pseudopotentials.}

\begin{tabular}{|c|c|c|c|}
\hline 
scheme&
\multicolumn{3}{c|}{\( E_{v} \) unrelaxed \( \left( eV\right)  \)}\\
&
LDA&
PW91&
PBE\\
\hline
Ultrasoft, MP q=2&
 4.068&
4.106&
4.113\\
\hline
Ultrasoft, MP q=4&
3.995&
4.018&
4.025\\
\hline
Normconserving, MP q=4&
4.016&
4.040&
4.051\\
\hline
\end{tabular}
\end{table}

\begin{table}

\caption{\label{table: final results}Final parameters for the fully converged
calculation of the silicon vacancy.}

\begin{tabular}{|c|c|}
\hline 
Quantity&
 Value\\
\hline
Number of atoms&
 256\\
\hline
Symmetry of supercell&
 BCC\\
\hline
Basis set size&
 160 eV\\
\hline
Brillouin zone sampling density&
 0.033 \AA\( ^{-1} \)\\
\hline
Vacancy formation energy (unrelaxed)&
 4.36 eV\\
\hline
Vacancy formation energy (relaxed)&
 3.17 eV\\
\hline
Symmetry of defect (unrelaxed)&
 \( T_{d} \)\\
\hline
Symmetry of defect (relaxed)&
 \( D_{2d} \)\\
\hline
Volume of defect (unrelaxed)&
 6.671 \AA\( ^{3} \)\\
\hline
Volume of defect (relaxed)&
4.874 \AA\( ^{3} \) \\
\hline
\end{tabular}
\end{table}

\clearpage

\end{document}